\documentclass[iop,useAMS,usenatbib]{emulateapj}

\usepackage{graphicx}
\usepackage{multirow}
\usepackage{color}
\usepackage{rotating}

\newcommand{\msun}{${\rm M_{\sun}}$}

\def\ltsima{$\; \buildrel < \over \sim \;$}
\def\simlt{\lower.5ex\hbox{\ltsima}}
\def\gtsima{$\; \buildrel > \over \sim \;$}
\def\simgt{\lower.5ex\hbox{\gtsima}}
\def\kms{{\rm\,km\,s^{-1}}}

\def\kpc{{\rm\,kpc}}

\def\msun{{\rm\,M_\odot}}
\def\lsun{{\rm\,L_\odot}}

\def\pc{{\rm\,pc}}

\newcommand{\fmmm}[1]{\mbox{$#1$}}
\newcommand{\scnd}{\mbox{\fmmm{''}\hskip-0.3em .}}
\newcommand{\scnp}{\mbox{\fmmm{''}}}
\newcommand{\mcnd}{\mbox{\fmmm{'}\hskip-0.3em .}}


\def\AA{$\; \buildrel \circ \over {\rm A}$}

\def\deg{^\circ}
\def\degg{\hbox{$\null^\circ$\hskip-3pt .}}
\def\sec{\hbox{"\hskip-3pt .}}

\def\s{\ifmmode \widetilde \else \~\fi}
\def\={\overline}

\def\spose#1{\hbox to 0pt{#1\hss}}

\def\lta{\mathrel{\spose{\lower 3pt\hbox{$\mathchar"218$}}
     \raise 2.0pt\hbox{$\mathchar"13C$}}}
\def\gta{\mathrel{\spose{\lower 3pt\hbox{$\mathchar"218$}}
     \raise 2.0pt\hbox{$\mathchar"13E$}}}
\def\Dt{\spose{\raise 1.5ex\hbox{\hskip3pt$\mathchar"201$}}}    
\def\dt{\spose{\raise 1.0ex\hbox{\hskip2pt$\mathchar"201$}}}    
\def\del{\nabla}

\def\dotsfill{\leaders\hbox to 1em{\hss.\hss}\hfill}
\def\sun{\odot}
\def\earth{\oplus}

\def\ltsima{$\; \buildrel < \over \sim \;$}
\def\gtsima{$\; \buildrel > \over \sim \;$}
\def\lsim{\lower.5ex\hbox{\ltsima}}
\def\gsim{\lower.5ex\hbox{\gtsima}}
\def\lapp{\ifmmode\stackrel{<}{_{\sim}}\else$\stackrel{<}{_{\sim}}$\fi}
\def\gapp{\ifmmode\stackrel{>}{_{\sim}}\else$\stackrel{<}{_{\sim}}$\fi}

\def\NMODY{{\sc n-mody}}
\def\gvext{\vec{g}_{ext}}
\def\ratilde{\tilde{r}_a}

\shorttitle{A kinematic survey of NGC 2419}
\shortauthors{Ibata et al.}

\begin{document}

\title{The globular cluster NGC~2419: a crucible for theories of gravity\altaffilmark{1}}

\author{R. Ibata\altaffilmark{2}, A. Sollima\altaffilmark{3}, C. Nipoti\altaffilmark{4}, M. Bellazzini\altaffilmark{5}, S.C. Chapman\altaffilmark{6}, E. Dalessandro\altaffilmark{4}}

\altaffiltext{1}{Some of the data presented herein were obtained at the W.M. Keck Observatory, which is operated as a scientific partnership among the California Institute of Technology, the University of California and the National Aeronautics and Space Administration. The Observatory was made possible by the generous financial support of the W.M. Keck Foundation. Also based on observations obtained with MegaPrime/MegaCam, a joint project of CFHT and
CEA/DAPNIA, at the Canada-France-Hawaii Telescope (CFHT) which is operated by the National Research Council (NRC) of Canada, the Institute National des Sciences de l'Univers of the Centre National de la Recherche Scientifique of France, and the University of Hawaii.}

\altaffiltext{2}{Observatoire Astronomique, Universit\'e de Strasbourg, CNRS, 11, rue de l'Universit\'e, F-67000 Strasbourg, France; rodrigo.ibata@astro.unistra.fr}

\altaffiltext{3}{INAF - Osservatorio Astronomico di Padova, vicolo
dell'Osservatorio 5, 35122, Padova, Italy}

\altaffiltext{4}{Dipartimento di Astronomia, Universit\`a degli Studi di Bologna, via Ranzani 1, I-40127 Bologna, Italy}

\altaffiltext{5}{INAF - Osservatorio Astronomico di Bologna, via Ranzani 1, 40127, Bologna, Italy}

\altaffiltext{6}{Institute of Astronomy, Madingley Road, Cambridge CB3 0HA, UK}

\begin{abstract}
We present the analysis of a kinematic data set of stars in the globular cluster NGC~2419, taken with the DEIMOS spectrograph at the Keck II telescope.  Combined with a reanalysis of deep Hubble Space Telescope and Subaru Telescope imaging data, which provide an accurate luminosity profile of the cluster, we investigate the validity of a large set of dynamical models of the system, which are checked for stability via N-body simulations. We find that isotropic models in either Newtonian or Modified Newtonian Dynamics (MOND) are ruled out with extremely high confidence. However, a simple Michie model in Newtonian gravity with anisotropic velocity dispersion provides an excellent representation of the luminosity profile and kinematics of the cluster. The anisotropy profiles of these models ensure an isotropic center to the cluster, which progresses to extreme radial anisotropy towards the outskirts. In contrast, with MOND we find that Michie models that reproduce the luminosity profile either over-predict the velocity dispersion on the outskirts of the cluster if the mass to light ratio is kept at astrophysically-motivated values, or else they under-predict the central velocity dispersion if the mass to light ratio is taken to be very small. We find that the best Michie model in MOND is a factor of $\sim10^4$ less likely than the Newtonian model that best fits the system. A likelihood ratio of 350 is found when we investigate more general models by solving the Jeans equation with a Markov-Chain Monte Carlo scheme. We verified with N-body simulations that these results are not significantly different when the MOND external field effect is accounted for. If the assumptions that the cluster is in dynamical equilibrium, spherical, not on a peculiar orbit, and possesses a single dynamical tracer population of constant $M/L$ are correct, we conclude that the present observations provide a very severe challenge for MOND.
\end{abstract}


\section{Introduction}
\label{sec:Introduction}

Are accelerations in the weak regime well described by the Newtonian approximation to General Relativity, or is a modification to this theory, such as the Modified Newtonian Dynamics (MOND; \citealt{Milgrom:1983p15031}), required? MOND was proposed 30 years ago as an alternative to dark matter to explain the approximately flat rotation curves in the outer parts of disk galaxies. In this theory the gravitational acceleration deviates significantly from that predicted by Newtonian gravity only in the vicinity or below a value $a_0 (\sim 1.2 \times 10^{-8}\, {\rm cm \, s^{-2}})$. One implementation of this idea, due to \citet{Bekenstein:1984p15032}, is to modify Poisson's equation in Newtonian gravity $\del^2 \varphi_N = 4 \pi G \rho$ to:
\begin{equation}
\label{eq_mond}
\vec{\del} \cdot \Big[ \mu\Big( {{| \vec\del \varphi |} \over {a_0}} \Big) \vec\del \varphi \Big] = 4 \pi G \rho \, ,
\end{equation}
where $\varphi$ is the MOND potential, and $\mu$ is an interpolating function that changes between $\mu(x)=1$ in the Newtonian regime, and $\mu(x)=x$ in the deep MOND limit of low accelerations. For a finite-mass isolated system the boundary conditions of equation~(\ref{eq_mond}) are $| \vec\del \varphi |\to 0$ for $|{\vec r}|\to \infty$, where ${\vec r}$ is the position vector.

While MOND was initially developed to explain the rotation curves of galaxies and the Tully-Fisher relation \citep[see, e.g.,][]{Tully:1977p17030, McGaugh:2005p15856}, it was not evident that it would hold up to closer scrutiny in other astrophysical systems. Indeed, many observational challenges to MOND have been presented over the years.  The theory has remained surprisingly resilient to this observational assault \citep[][]{Sanders:2002p15425}, though in some cases there are difficulties in reconciling the observations with the predictions of MOND.  These include problems with the growth of cosmological perturbations \citep{Dodelson:2006p17084}, the offset between lensing mass and baryons in the Bullet Cluster \citep{Clowe:2006p15213}, Solar System tests \citep{Milgrom:1983p15031,Sereno:2006p15936}, dynamical friction in dwarf galaxies \citep{Ciotti:2004p17054, SanchezSalcedo:2006p17060, Nipoti:2008p17050, Angus:2009p17066}, and the kinematics and density profile of satellite galaxies \citep{Klypin:2009p17870}, to list a few.  However the main-stream explanations of these systems that involve dark matter are not themselves without problems, especially at the galactic scale where the discrepancy between the predictions of standard Cold Dark Matter theory ($\Lambda$CDM) and observations appears to be greatest.  The issue of the missing satellite galaxies, the possible non-existence of dark matter cusps and the high angular momentum of galactic disks are among the more famous problems that pose difficulties to $\Lambda$CDM \citep{Binney:2004p17034, Primack:2009p17036}, although it is possible that the discrepancies are due to our incomplete understanding of the physics of baryonic matter. It is therefore very important to challenge these theories of gravity in the most stringent ways we have access to, using well-understood astrophysical systems, that can be measured and probed with great accuracy.

It has been appreciated for a number of years \citep{Scarpa:2003p17001, Haghi:2009p16438} that the dynamics of globular clusters may give a means to assess the fundamental question above, since the accelerations that stars experience on the periphery of these systems are exceedingly low. Indeed, if one were to consider a globular cluster in isolation, one would expect to see a drop in the accelerations that stars experience down to values approaching zero on the periphery of the cluster. In reality, however, these systems reside inside a host galaxy, so in general they cannot be treated as isolated. In Newtonian dynamics, provided that the gravitational field of the galaxy can be considered uniform across the extent of the cluster (i.e. if tidal effects are negligible), the internal dynamics of the cluster are the same as if the cluster were isolated. This is not the case in MOND, in which, due to the non-linearity of the theory, the internal dynamics of a stellar system is influenced even by a uniform external field \citep[in other words, the strong equivalence principle is not valid;][]{Milgrom:1983p15031}. In the considered case of MOND as modified gravity, this external field effect (EFE) can be accounted for in the boundary conditions of Equation~(\ref{eq_mond}), which are $\vec\del \varphi \to \vec \del \varphi_{ext}$ for $|{\vec r}|\to \infty$ in the presence of a uniform external field $\gvext=- \vec \del \varphi_{ext}$ \citep{Bekenstein:1984p15032}. An important consequence of the EFE is that the internal dynamics of a MOND system with even low internal accelerations becomes almost Newtonian in the presence of a sufficiently strong external field. It is then clear that for a globular cluster to be useful in the attempt to distinguish between Newtonian gravity and MOND, not only must its internal accelerations be low, but also the external galactic field it experiences must be low, compared to $a_0$.

\begin{figure}
\begin{center}
\includegraphics[angle=0, bb= 60 80 560 750, clip, width=\hsize]{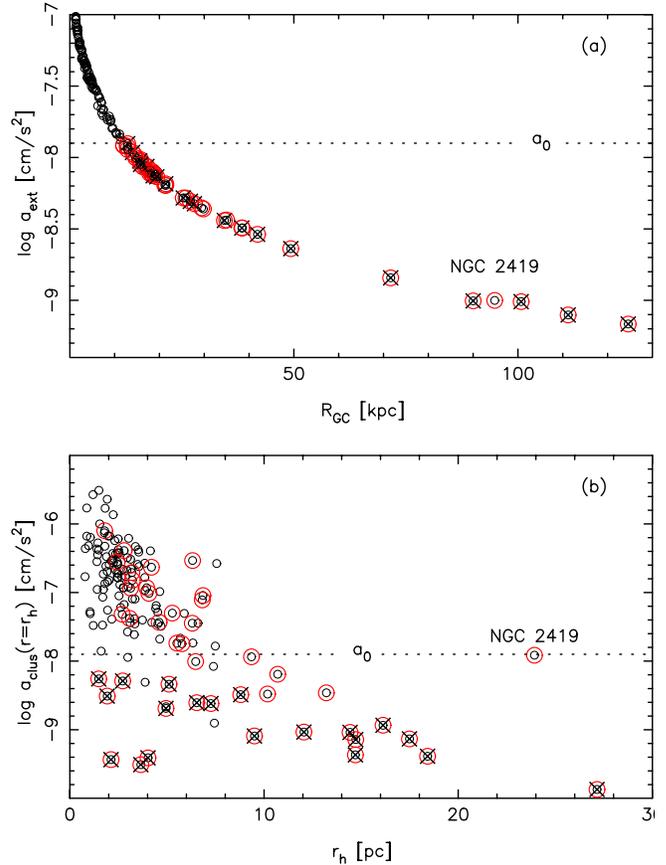}
\end{center}
\caption{Top panel: The expected external accelerations that Galactic globular clusters experience as a function of their radial distance from the center of the Milky Way. The realistic Galactic mass model from \citet{Dehnen:1998p10420} is assumed, and the actual three-dimensional positions of the clusters are used. Though the acceleration is computed for a Newtonian model of the Milky Way with dark matter, we expect similar values in MOND if MOND has to reproduce the rotation curve of the Milky Way. Those globular clusters that experience an acceleration below $a_0$ are circled red, and the subset of these clusters that are expected to contain fewer than 50 RGB stars are further marked with crosses. The crossed-out clusters are poor targets since they contain very few stars amenable to spectroscopic follow-up \citep[see the discussion in][]{Haghi:2009p16438}. Clearly, NGC~2419 is the only rich cluster in the outer Galaxy. (We have shifted the two clusters on either side of NGC~2419 --- Eridanus and Palomar~3 --- by $5\kpc$ to avoid a confusing overlap at $R_{GC} \sim 95\kpc$). Bottom panel: the expected internal acceleration at the half-light radius, assuming Newtonian gravity, for the same sample of globular clusters, as derived from the compilation by \citet{Harris:1996p11754}. Clearly NGC~2419, the target of this study, is minimally affected by the external Galactic acceleration field, and is also sufficiently extended so as to have a significant amount of its member stars in the low acceleration regime. The MOND $a_0$ level is indicated with a dotted line in both panels.}
\label{fig:GCs_in_MW}
\end{figure}

Figure~\ref{fig:GCs_in_MW}a illustrates the external gravitational acceleration due to the Milky Way that members of the Galactic globular cluster population experience. For the majority of the clusters, the external field accelerations are substantially above the MOND characteristic acceleration $a_0$. Note also that the Galactocentric distances shown in Figure~\ref{fig:GCs_in_MW}a refer to the present-day orbital phase, so that even some of the more distant clusters will pass closer to the Milky Way center during their orbits. It is only the most distant clusters, well in the realm of the distant halo, that can be considered to be sufficiently isolated to probe the low acceleration regime. Also relevant is the fraction of the cluster stars that lie within the MOND regime: this fraction is substantial only for the most extended clusters (Figure~\ref{fig:GCs_in_MW}b).

\begin{figure}
\begin{center}
\includegraphics[bb= 30 90 560 740, angle=270, clip, width=\hsize]{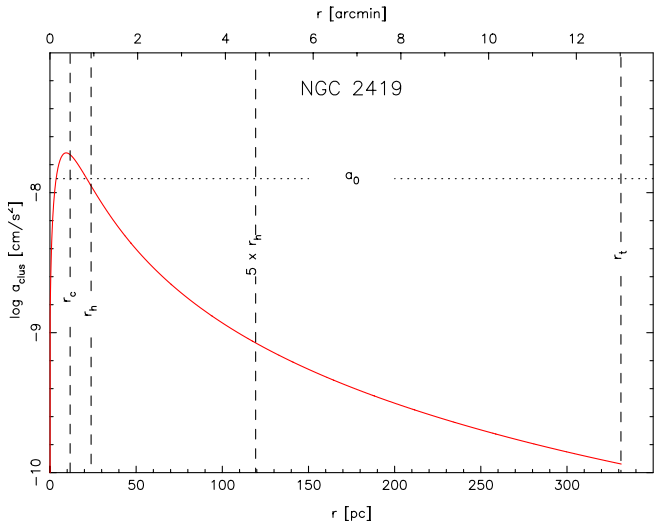}
\end{center}
\caption{The Newtonian internal accelerations in the best-fitting Michie model of NGC~2419 found in this study (see \S4), as a function of distance from the cluster center. The core radius $r_c$, the half-light radius $r_h$, and the tidal radius $r_t$ are marked. With our improved parameter constraints, it is evident that even at the cluster half-light radius, the acceleration that stars experience is below the $a_0$ value.}
\label{fig:acc_in_GC}
\end{figure}

NGC~2419 is a particularly promising globular cluster to study in this context.  First, as indicated in the top panel of Figure~\ref{fig:GCs_in_MW}, it is one of the most distant objects among the known population of Milky Way clusters, so it experiences a low external force. Indeed, the acceleration due to the Galaxy at that location, according to the Milky Way model of \citet{Dehnen:1998p10420}, is well below the critical MOND acceleration value ($g_{ext}\sim0.1~a_0$). Furthermore, NGC~2419 is a very extended cluster, which implies that the accelerations on its constituents are, on average, very low.  Thus is illustrated in the lower panel of Figure~\ref{fig:GCs_in_MW}, where the internal acceleration is given as a function of the half light radius. Clearly the expected accelerations drop rapidly further out in the cluster, and as we show in Figure~\ref{fig:acc_in_GC}, slightly beyond the half mass radius in NGC~2419 fall well below $a_0$. The high luminosity of this cluster (which is the fourth most luminous in the Milky Way, after $\omega$-Cen, M54 and NGC~6388) also means that there are large numbers of target stars for study, even in the shortest evolutionary phases, such as the the upper Red Giant Branch (RGB), and also ensures that the velocity dispersion is large enough to resolve easily with current instrumentation. These properties make NGC~2419 a much more favourable subject than nearby massive clusters \citep{Lane:2010p16975}, or distant low-mass clusters like Palomar~14 \citep{Jordi:2009p16532, Gentile:2010p16618} that have previously been scrutinized to test MOND.

Throughout this study, unless otherwise indicated, we adopt the distance measurement of $(m-M)_0=19.71 \pm 0.08$ \citep{DiCriscienzo:2011p15907}; this implies a distance of $87.5\kpc$ and an angular scale of $25.452\pc$ per arcmin. We assume the center of the cluster is located at $07^h 38^m 08^s.47 +38\deg 52' 55\sec0$ \citep[][hereafter D08]{Dalessandro:2008p15030}, which is towards the Galactic anti-center, where the reddening is measured to be $E({\rm B-V}) = 0.08 \pm 0.01$ \citep{Ripepi:2007p15041, DiCriscienzo:2011p15907} from RR~Lyrae lightcurves.

\begin{table}
\begin{center}
\caption{Properties of the globular cluster NGC~2419.}
\label{tab:dispersions}
\begin{tabular}{ccc}
\tableline\tableline
Parameter & value & source \\
\tableline
RA & $07^h 38^m 08^s.47$ & 1\\
Dec & $+38\deg 52' 55\sec0$ & 1 \\
$\ell$ & $180.3696$ & \\
$b$   & $+25.2416$ & \\
$E({\rm B-V})$ & $0.08 \pm 0.01$~mag & 2 \\
$(m-M)_0$ & $19.71 \pm 0.08$ & 2 \\
Distance & $87.5\pm 3.3\kpc$ & 2 \\
Angular scale & $25.452\pc$ per arcmin & \\
${\rm [Fe/H]}$ & -2.1 (Zinn \& West scale) & 2 \\
$\mu_V(0)$ & 19.55 & 3 \\
$V_t$ & $10.47\pm0.07$ & 2 \\
${M/L}_V$ & $1.90\pm0.16 \msun/\lsun$ & 4 \\
Mass & $ 9.12\times10^5$ & 5 \\
\tableline\tableline
\end{tabular}
\tablecomments{The sources are: 1 = \citet{Dalessandro:2008p15030}, 2 = \citet{DiCriscienzo:2011p15907}, 3 = \citet{Bellazzini:2007p14801}, 4 = \citet{McLaughlin:2005p15121}, 5 = this study. Rows without source information are derived from other table parameters.}
\end{center}
\end{table}

\section{Observations}
\label{sec:Observations}

The analysis in this article is based upon several photometric and spectroscopic data sets. The main photometric data set, which serves to determine the stellar surface brightness profile of the cluster, is derived from images taken with the Advanced Camera for Surveys (ACS) on board the Hubble Space Telescope and much wider-field images taken with the Suprime-Cam camera on the Subaru Telescope, as derived by D08. During the course of the analysis presented below, we realized that D08 had overestimated the uncertainties on the density profile, making the model comparisons with the data highly surprising in the sense that they were ``too good to be true''. To address this problem we used their photometric catalog to re-determine the light profile of the cluster. We define a sub-sample of 7849 stars, by selecting from a color-magnitude box that incorporates the upper main sequence, the RGB and blue horizontal branch (BHB) populations, and limited the sample to ${\rm V = 23.5}$ (as argued by D08) to avoid incompleteness, and radial variations thereof.

\begin{figure}
\begin{center}
\includegraphics[bb= 50 90 560 740, angle=270, clip, width=\hsize]{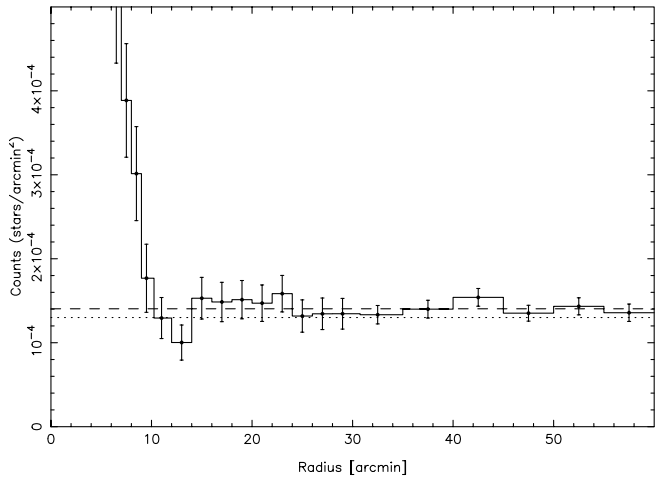}
\end{center}
\caption{The star-count profile in the CFHT field used to probe the background profile at large distance from the cluster. The stellar selection was chosen to be as close as possible to that of the main Suprime-Cam/ACS survey, and indeed the weighted mean background level of the CFHT counts between $20\arcmin$ and $60\arcmin$ is $0.00014 \, {\rm arcsec}^{-2}$ (dashed line), very similar to the Suprime-Cam average of $0.00013 \pm 0.00002 \, {\rm arcsec}^{-2}$ (dotted line) at $r > 612\scnp$. The CFHT data shows that there is no significant contribution from the cluster at radii beyond $\sim 10\arcmin$.} 
\label{fig:CFHT_profile}
\end{figure}

Star counts were obtained in 15 radial annuli (or portions of radial annuli) within $r < 612\scnp$, the largest circle that is fully included in the Suprime-Cam field. As can be appreciated from Figure~2 of D08, although the Subaru data cover the cluster well, they do not extend sufficiently far away from the cluster so as to provide a background comparison region that is unambiguously free of cluster member stars. Accurate background subtraction is a crucial issue for our study, so to explore better the transition between the cluster and the background population\footnote{The ``background'' population in reality is of course mostly due to foreground Galactic stars.}, we obtained archival images of the ROSAT cluster RXCJ0736.4+3925 taken with the MegaCam Camera on the Canada-France-Hawaii telescope (CFHT). With exposures of 2160~sec in g and 1920~sec in the r-band, these images cover a $1\deg\times1\deg$ field, with NGC~2419 located on the southern edge of the mosaic. The MegaCam images were processed and reduced to photometric catalogs in an identical way to that described in \citet{Ibata:2007p160}, using the Cambridge Astronomical Survey Unit pipeline \citep{Irwin:2001p18115}. By cross-identifying the sources in the CFHT and Subaru catalogs, we devised a color-magnitude selection in the g and r bands that was as close as possible to that made in the V and I bands from the D08 sample. Figure~\ref{fig:CFHT_profile} shows the star-counts profile of the point sources in the CFHT image that lie within the defined stellar locus. Clearly, beyond $10\arcmin$ and up to $1\deg$ from the cluster center the star counts are consistent with a flat background. This justifies our choice of using the area at $r>612\scnp$ in the Suprime-Cam/ACS survey (our main photometric catalog) to estimate the background count level ($0.00013 \pm 0.00002 \, {\rm arcsec}^{-2}$). 

The absolute photometric zero-point of the Suprime-Cam/ACS profile was defined by matching the data at $ r \le 16\scnd5$ with the 6 innermost points of the profile by \citet[][hereafter B07]{Bellazzini:2007p14801} that are based on aperture photometry of the ACS image. The gaps between detectors and the most obvious holes due to saturated stars were taken into account in computing the area of the annuli. The resulting profile is tabulated in Table~\ref{tab:starcounts}; apart from the improved uncertainty estimates, this profile is in agreement with that of D08.  Note, however, that the non-azimuthally symmetric geometry of the missing areas (see Figures 1 \& 2 of D08) gives rise to slight radial biasses in the star-counts profile, which must be accounted for in any fitting or modeling of the cluster by the use of appropriate window functions\footnote{The window functions and color-magnitude selection functions are available from the authors, on request.}.

\begin{table*}
\begin{center}
\caption{The star-counts profile of NGC~2419.}
\label{tab:starcounts}
\begin{tabular}{ccccrrrccccc}
\tableline\tableline
$r_i$  &$r_o$  &$a_{min}$  &$a_{max}$  &   $r_m$ \,  &  $\langle r \rangle$  \, &$\sigma_r$ \, &  $\mu_V$  & $\delta\mu_V$  & $N_{tot}$&     $A_{eff}$  &source\\
($\scnp$) &($\scnp$) & ($\deg$) & ($\deg$) & ($\scnp$) \, &($\scnp$) \, &($\scnp$) \, & ${\rm mag/arcsec^{2}}$& ${\rm mag/arcsec^{2}}$ &   & ${\rm arcsec^2}$ & \\
\tableline
   0 &   5 & -180 &    0 &   2.50 &  3.202 &  1.156 & 19.652 & 0.097 &   124 & $3.93\times10^1$ & 1\\
   5 &  10 & -180 &    0 &   7.50 &  7.735 &  1.433 & 19.857 & 0.062 &   308 & $1.17\times10^2$ & 1\\
  10 &  15 & -180 &    0 &  12.50 & 12.572 &  1.384 & 20.019 & 0.053 &   415 & $1.84\times10^2$ & 1\\
  15 &  20 & -180 &    0 &  17.50 & 17.495 &  1.432 & 20.233 & 0.048 &   508 & $2.75\times10^2$ & 1\\
  20 &  30 & -180 &    0 &  25.00 & 24.822 &  2.861 & 20.740 & 0.036 &   910 & $7.85\times10^2$ & 1\\
  30 &  40 & -130 &    0 &  35.00 & 34.924 &  2.902 & 21.176 & 0.044 &   616 & $7.94\times10^2$ & 1\\
  40 &  65 & -100 &    0 &  52.50 & 50.962 &  7.153 & 22.159 & 0.040 &   719 & $2.29\times10^3$ & 1\\
  65 &  80 & -180 &  180 &  72.50 & 71.989 &  4.446 & 23.090 & 0.036 &   910 & $6.83\times10^3$ & 2\\
  80 & 100 & -180 &  180 &  90.00 & 89.259 &  5.843 & 23.740 & 0.038 &   828 & $1.13\times10^4$ & 2\\
 100 & 130 & -180 &  180 & 115.00 &113.743 &  8.611 & 24.475 & 0.038 &   808 & $2.17\times10^4$ & 2\\
 130 & 180 & -180 &  180 & 155.00 &152.306 & 14.418 & 25.365 & 0.039 &   790 & $4.79\times10^4$ & 2\\
 180 & 260 & -180 &  180 & 220.00 &212.092 & 23.296 & 26.660 & 0.057 &   386 & $7.58\times10^4$ & 3\\
 260 & 360 & -180 &  180 & 310.00 &304.503 & 29.046 & 27.879 & 0.069 &   288 & $1.65\times10^5$ & 3\\
 360 & 460 & -180 &  180 & 410.00 &402.642 & 27.708 & 29.250 & 0.123 &   135 & $2.31\times10^5$ & 3\\
 460 & 612 & -180 &  180 & 536.00 &528.313 & 44.745 & 30.991 & 0.292 &   104 & $4.74\times10^5$ & 3\\
\tableline\tableline
\end{tabular}
\tablecomments{The quantities $r_i$ and $r_o$ are the inner and outer radii of the annulus of effective area $A_{eff}$ that has mid-value $r_m[=0.5(r_i+r_o)]$ and that is composed of $N_{tot}$ stars at an average radius $\langle r \rangle$ with standard deviation $\sigma_r$. The surface brightness, not corrected for extinction, is $\mu_V$, with Poisson uncertainties $\delta\mu_V$. The ACS survey does not fully cover the inner region of the cluster (see Fig.~1 of D08), so to avoid radial biasses, we construct the profile from data in an angular range $a_{min}$ to $a_{max}$ from $-180\deg$ to $180\deg$ (where West $=0\deg$ and North $=+90\deg$). The final column lists the source of the data, with 1$=$ACS, 2$=$ACS$+$Suprime-Cam, 3$=$Suprime-Cam.}
\end{center}
\end{table*}

Targets for follow-up spectroscopy were selected from an earlier compilation of photometric data constructed by B07, which is a combination of photometry measured from the same ACS images, as well as from ground-based images which cover a wider field of view. The ground-based photometry includes the WIYN study by \citet{Saha:2005p9610}, Sloan Digital Sky Survey (SDSS) data \citet{AdelmanMcCarthy:2006p14983} as well as images retrieved from the archive of the Telescopio Nazionale Galileo (TNG).

\begin{figure*}
\begin{center}
\includegraphics[angle=0, bb= 75 20 560 777, clip, width=14cm]{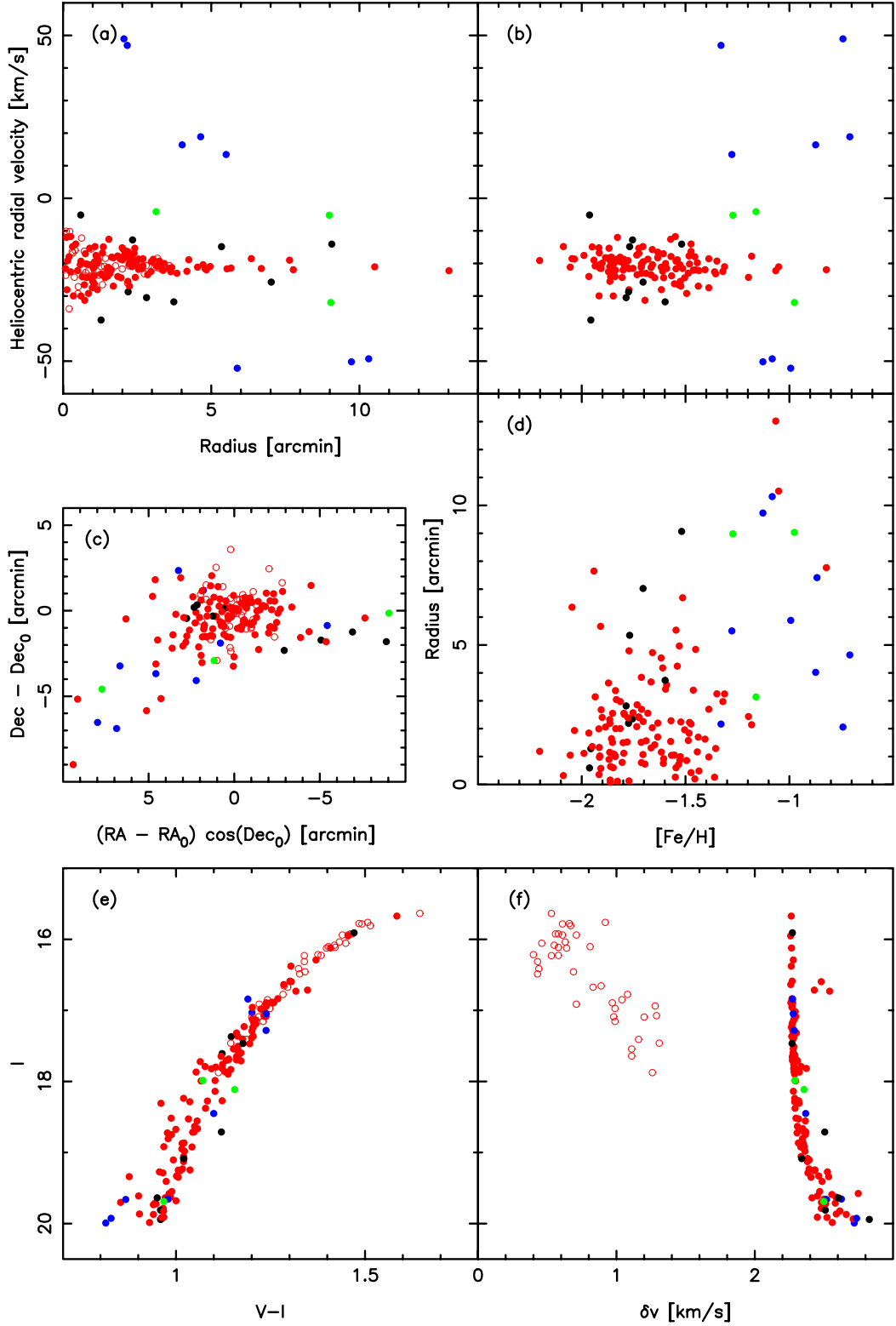}
\end{center}
\caption{Parameter projections of the kinematic data set. The B09 sample is shown with open circles, while filled dots are from the present DEIMOS survey.  The stars are marked either red, black, green, or blue according to their likely cluster membership. Blue points are very likely contaminants, as they lie more than $20\kms$ from the cluster mean. Red points mark those stars that are inside 2 times the velocity dispersion of the MOND model shown in Figure~\ref{fig:vel_radius}, and constitute the cleanest sample. Black and green points lie beyond that model, but inside $20\kms$ from the cluster mean, with the black points having ${\rm [Fe/H] < -1.3}$ and the green ${\rm [Fe/H] > -1.3}$. Note that B09 do not provide metallicity information, so open circles are absent from panels `b' and `d'.}
\label{fig:data}
\end{figure*}

The main spectroscopic observations presented here were taken with the DEIMOS multi-object slit spectrograph \citep{Faber:2003p14992} mounted on the Keck II telescope at Mauna Kea, Hawaii on 2008 October 27th. We employed the high-resolution 1200~line/mm grating coupled with the OG550 order-blocking filter to give access to the spectral range $\sim 6500$ -- $9500$\AA\ at a dispersion of 0.33\AA$/{\rm pixel}$ (i.e. $12\kms/{\rm pixel}$ at 8500\AA). The spectrograph has a scale of $0\scnd1185/{\rm pixel}$ and an anamorphic magnification of $0.7$. The spectrograph camera has two rows of detectors `blue' and `red'; in order to maintain the \ion{Ca}{2} triplet lines on the red side of the camera to avoid the inter-CCD gap, we set the central wavelength to $8000$\AA. Narrow slits of $0\scnd7$ were milled to optimize the spectral resolution, which was determined to be $\sim 1.2$\AA, based upon measurements of the width of arc lamp lines. Target stars were selected from the B07 catalogue with ${\rm I < 20}$ and within a narrow color-magnitude region around the cluster Red Giant Branch (RGB), as shown in Figure~\ref{fig:data}e. Two masks milled with slitlets for 85 and 77 targets, were observed with exposure times of $3\times 1200~{\rm s}$ and $3\times1000~{\rm s}$, respectively (we will henceforth refer to these as masks 1 and 2). The weather was clear during these exposures and the seeing was approximately $0\scnd7$, fortuitously matching the slit widths.

\begin{figure}
\begin{center}
\includegraphics[angle=0, bb= 60 80 560 740, clip, width=\hsize]{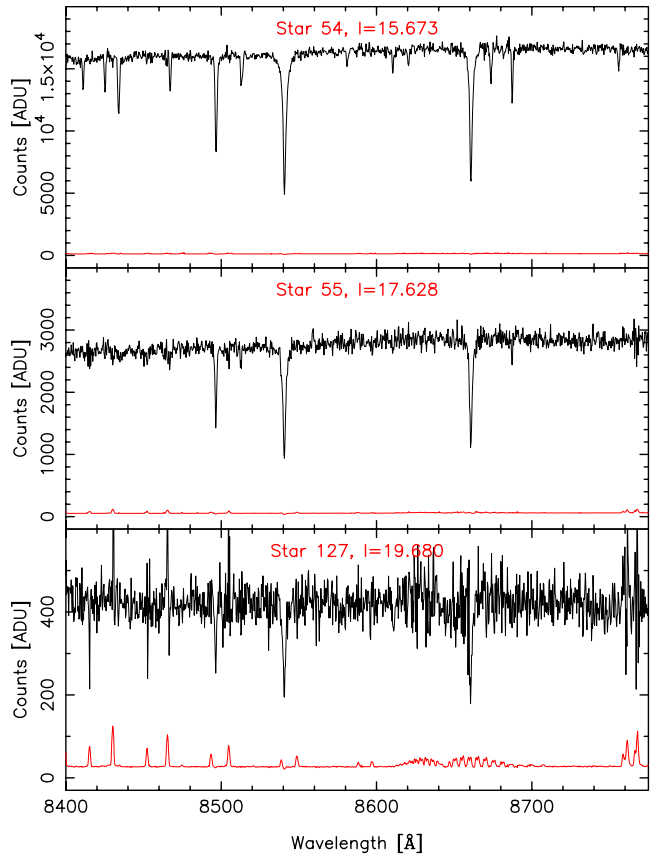}
\end{center}
\caption{Spectra representative of bright, medium and faint stars in the spectroscopic sample that are confirmed members of NGC~2419. The wavelength range shown highlights the main region of interest around the prominent \ion{Ca}{2} triplet lines at $8498.02$, $8542.09$, and $8662.14$\AA. The error spectrum is also shown (red line) in each panel, as well as the identification number of the star in Table~\ref{tab:data}. Note that the pipeline discussed in the text produces non-binned versions of the spectra (which are just pixel fluxes with their corresponding wavelengths); here such data have been binned up for display purposes only.}
\label{fig:spectra}
\end{figure}

In order to convey the quality of the data, we display in Figure~\ref{fig:spectra} the spectra (and the corresponding noise spectra) for three stars that are representative of the brightest, of intermediate and of the faintest stars in the cluster sample. The signal to noise of the spectra are clearly high, exceeding $S/N=10$ per pixel for the faintest stars in the sample and reaching as high as $S/N \sim 70$ for the very brightest stars.

A new spectral reduction package was constructed, partly for this project, in an attempt to improve object extractions, wavelength calibration and sky subtraction, as well as to provide more reliable uncertainty estimates. The procedure will be presented in detail in a forthcoming contribution (Ibata et al., in preparation); a brief summary is sufficient for the purpose of the present work. The aim of the software is to avoid any resampling of the observed two-dimensional spectroscopic frames: each observed pixel and its corresponding uncertainty is kept all the way to the final calibrated product, at which point the spectra may be trivially collapsed into one-dimensional form for visualization purposes, if necessary. The reduction steps are as follows:
\begin{itemize}
\item 
identify cosmic rays in the images if multiple exposures are available.
\item
correct for scattered light.
\item
perform a flat-field correcting for pixel-to-pixel variations.
\item
perform an illumination and the slit function correction.
\item
correct for fringes.
\item
wavelength-calibrate every pixel, using the arc-lamp with a correction from sky-lines.
\item
perform a two-dimensional sky-subtraction (see \citealt{Kelson:2003p14171}).
\item
extract spectra without resampling in a small spatial region around the target.
\end{itemize}
Thus the final product of the pipeline is a large set of pixel fluxes for every target. Each of these pixels has an associated flux uncertainty, a wavelength, a wavelength uncertainty, and the value of the target spatial profile at the pixel.

It is then possible to measure the velocity of the target using a Bayesian approach. We adopt a simple model for a normalized stellar spectrum that consists of a continuum broken by three Gaussian absorption profiles located at the rest wavelengths of the three prominent Calcium II triplet lines (8498.02\AA, 8542.09\AA, 8662.14\AA). The stellar continuum is estimated using a robust algorithm that iteratively filters out absorption and emission lines and computes a running median of the spectrum. We multiply the normalized model by the continuum function to compare to the observations. A Markov-Chain Monte-Carlo algorithm then finds the best Doppler shift and Ca II line widths given the individual pixel measurements, as well as the uncertainties on these parameters. Due to the high signal to noise of these observations and the excellent wavelength solution of the arc-lamp reductions, the velocity errors determined in this way for our NGC~2419 targets are very low, typically below $1\kms$. This can be achieved, despite the spectral resolution of $\sim 42\kms$, because the $S/N$ is sufficient to provide good centering constraints on the absorption lines.

However, unlike fibre-fed spectrographs, slit spectrographs have the the disadvantage that small milling errors (or large astrometry errors) will cause a target to be offset with respect to the slit, giving rise to an artificial velocity offset if the effect is not corrected. Given the geometry of DEIMOS, if a slit is erroneously milled $0\scnd1$ displaced along the dispersion direction, one would naively expect a $14\kms$ velocity shift; in practice though, the tolerance is less strict than this estimate suggests, since atmospheric turbulence smears out the incoming light. Indeed in poor seeing conditions such errors will be minimized. Another way that this problem can arise is from misalignment of the mask. DEIMOS masks are generally aligned by centering bright stars (typically 4) in small boxes that have been milled into the mask; before starting an exposure the observer iterates on positional and rotational corrections until the stars are centered to the achievable accuracy, which is about $0\scnd05$. An error in the mask alignment along the dispersion direction shows up easily as a systematic offset in the OH sky-line wavelength, while a rotational error causes a radius-dependent shift, as would be expected. These misalignments are corrected for in the pipeline to high accuracy (better than $1\kms$).

Nevertheless, the milling and astrometry errors are unavoidable, and it is difficult to estimate their effect. One way to alleviate the problem is to make use of telluric absorption lines: the position of the lines is not dependent on the slit error, so they give a promising means to calibrate velocities (as argued by \citealt{Simon:2007p14174}). Unfortunately, the telluric features are broad and time-dependent; from inspecting the effect on repeat measurements we found that the corrections that we derived were actually adding noise. We decided therefore not to include a telluric absorption correction.

The two observed masks have 12 stars in common, and the r.m.s. scatter between velocity differences is $1.91\kms$. This is encouraging, as it means that the individual measurements are a factor of $\sim \sqrt{2}$ smaller.

Due to the positioning errors, it is difficult for DEIMOS to provide us an accurate absolute velocity. To circumvent this problem, we adopt the approach of bringing our measurements onto the zero-point defined by the High-Resolution Echelle Spectrometer (HIRES) study of \citet[][hereafter B09]{Baumgardt:2009p14321}. There are 7 stars in common between the DEIMOS sample and the HIRES sample (39 stars), we find an average offset of $\Delta{\overline{v}} = \overline{v}_{HIRES}-\overline{v}_{DEIMOS} = - 3.06\kms$ with an r.m.s. scatter of $2.37\kms$. This r.m.s. scatter is produced by a combination of the HIRES error on each star quoted by B09 ($\delta_{i,HIRES}$), the error on the DEIMOS velocities returned by our pipeline ($\delta_{i,DEIMOS}$, itself a combination of the wavelength calibration uncertainty and the pixel measurement uncertainties), and the milling/astrometry error, which we will refer to as $\delta_{ast}$. We make the plausible assumption that $\delta_{ast}$ is independent of magnitude. The condition:
$${{1}\over{n}} \sum_{i=1}^n {{  (v_{i,HIRES}-v_{i,DEIMOS} - \Delta \overline{v})^2}\over{\delta_{i,HIRES}^2+\delta_{i,DEIMOS}^2+\delta_{ast}^2}}=1 \, ,$$
where we are summing over the $n=7$ stars in common between the two samples, requires that the milling/astrometric error must be $\delta_{ast}=2.25\kms$.  We add this additional uncertainty in quadrature to our pipeline velocity uncertainty estimates. Note however, that HIRES, being a slit spectrograph, also suffers from the same centering uncertainty. However, given its higher spectral resolution the corresponding velocity offset errors should be lower, and we take them to be negligible compared to the DEIMOS $\delta_{ast}$.

In addition to the velocities, we also measure the Na~I equivalent width (which can be a useful population discriminant), and Ca~II line strengths to derive a metallicity in an identical way to that discussed in \citet{Ibata:2005p216}. The metallicity (on the \citealt{Carretta:1997p17874} scale) is calculated as:
${\rm [Fe/H]} = -2.66 + 0.42 [\Sigma  Ca  -  0.64 ({\rm V}_{HB}  - {\rm V})]$,
with
$\Sigma Ca  =  0.5 EW_{\lambda 8498}  + 1.0 EW_{\lambda  8542} + 0.6  EW_{\lambda 8662}$,
$({\rm V}_{HB}  - {\rm V})$ being a surface gravity correction relative to the V-magnitude of the horizontal branch. We adopt the value  of ${\rm V}_{HB}=20.45$ \citep{Harris:1996p11754}. Table~\ref{tab:data} lists the star number, the mask number (or B09 identification number), the position, the radial velocity measurement and corresponding uncertainty, the I-band magnitude and ${\rm V-I}$ color, the metallicity, and the projected radial distance $R$ from the cluster center. Those stars with multiple observations are marked in the final column.

\begin{figure}
\begin{center}
\includegraphics[angle=0, bb= 60 80 560 740, clip, width=\hsize]{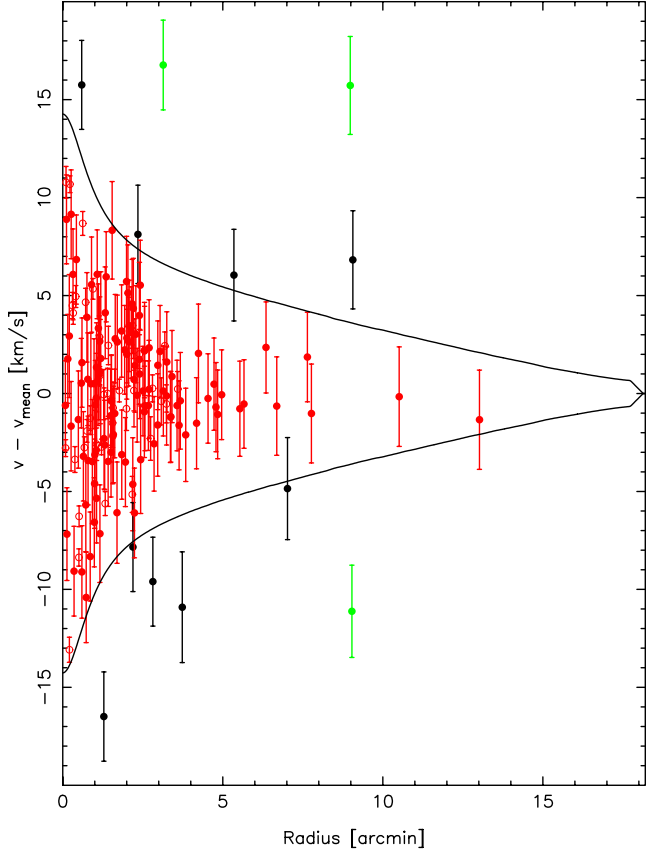}
\end{center}
\caption{Closeup of the cluster region of Figure~\ref{fig:data}a, showing velocity uncertainties. The graph markers have the same coding as in that figure. The line shows $2\times\sigma(r)$ according to the anisotropic $(M/L)_V=1.346$ MOND model (model \#24 in Table~\ref{tab:models}). Note that there are no plausible cluster members outside of this $\pm 20\kms$ velocity window (see Figure~\ref{fig:data}a).}
\label{fig:vel_radius}
\end{figure}

\begin{figure}
\begin{center}
\includegraphics[angle=0, bb= 60 80 560 740, clip, width=\hsize]{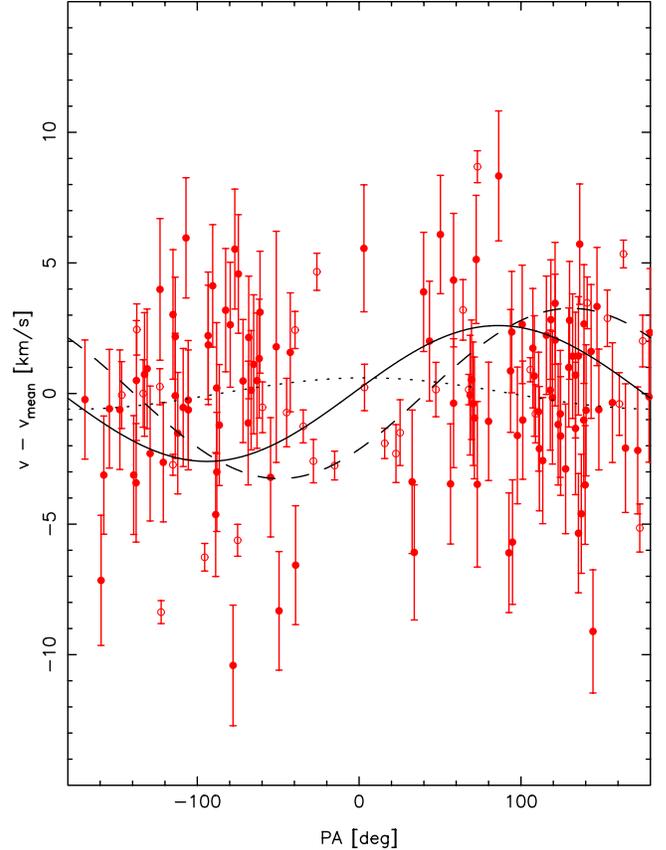}
\end{center}
\caption{Radial velocity as a function of position angle East of North for sample `A', but removing stars in the central $0\mcnd5$. The dashed curve shows the B09 fit to their sample. In contrast, fitting the data shown here gives the continuous curve, which has a slightly lower amplitude of $2.6\kms$, and has a 5\% chance of being simply a spurious result due to random sampling. However, fitting on the DEIMOS stars alone gives the dotted fit, which is certainly not significant.}
\label{fig:rotation}
\end{figure}

\subsection{Defining the kinematic sample}
\label{sec:sample}

The final sample comprises a total of 178 unique stars, of which 39 are taken from the B09 catalog. For those stars with multiple observations, we choose the measurement with the lowest estimated velocity uncertainty. Various parameter projections of the sample are shown in Figure~\ref{fig:data}: velocity as a function of radius, velocity as a function of metallicity, position on the sky, the metallicity-radius relation, color-magnitude position, and finally the velocity uncertainty-magnitude relation. While the majority of the observed stars are clumped close to the expected mean Heliocentric velocity of the cluster (Fig.~\ref{fig:data}a), there are clearly Galactic interlopers in the sample. The most obvious interlopers are marked blue and have velocity differences $>20\kms$ with respect to the cluster mean. In an attempt to make an initial selection to help guide the discussion, we take the velocity dispersion profile $\sigma_v(r)$ of the $(M/L)_V=1.346$ anisotropic MOND model (\#24) discussed in Section~\ref{sec:models} below. This model has the highest likelihood of all the MOND models we consider (for sample A), and its adoption in this discussion should favor the case for the MOND theory. Selecting stars within $2\times\sigma_v(r)$ allows us to make the initial separation between very likely cluster members (red) and the kinematic outliers (black and green) in Fig.~\ref{fig:data}a. Inspection of Figure~\ref{fig:data}b suggests that a metallicity cut at ${\rm [Fe/H]} = -1.3$ provides a means to sort the kinematic outliers into metal-poor (black) and metal-rich (green) samples\footnote{Recently, \citet{Cohen:2010p16255} have reported a significant metallicity spread in NGC~2419, a property we confirm in Figure~\ref{fig:data}d. Our metallicity distribution appears wider however, probably due to the inclusion of lower S/N spectra from stars at the faint end of the color-magnitude diagram.}. Unfortunately the other measured parameters do not provide us with strong additional information that could help discriminate between cluster members and interlopers (our Na~I line strength cut $\sum EW_{NaI} < 2.0$ rejects one dwarf star that would otherwise have been rejected anyway by velocity).

Due to its importance, in Figure~\ref{fig:vel_radius} we reproduce the velocity-radius relation, zooming-in on the cluster region, and displaying the velocity uncertainties. Note that we have refined the cluster mean velocity quoted by B09 ($-20.3\pm0.7\kms$), using instead a weighted mean of their data that gives $v_{mean}=-20.9\kms$; with this value the DEIMOS velocity distribution is noticeably more symmetric. The lines delineate $2\times\sigma(r)$ of the MOND model mentioned previously, a limit chosen to incorporate all of the sample of B09. Inspection of Figure~\ref{fig:data} shows that there are no stars with ${\rm [Fe/H] < -1.3}$ that have velocities beyond $20\kms$ of the cluster mean, which in turn strongly suggests all stars with ${\rm [Fe/H] < -1.3}$ are cluster members. This includes all stars marked in black in Figure~\ref{fig:vel_radius}, many of which have error bars that overlap the $|v(r)| < 2\times\sigma(r)$ region. Nevertheless, several of the black data points lie well outside of that region and, at face value, greatly enhance the local velocity dispersion.  We will return to this issue in Section~\ref{sec:analysis}. The three green points at $(R,v-v_{mean})=(3\mcnd1, 16.8\kms), (9\mcnd0, 15.7\kms), (9\mcnd0, -11.1\kms)$, are most probably Galactic interlopers, and can be safely removed from the sample. Of the four red points at $R>7\arcmin$, it is interesting that three are metal rich, having $-1.07 \le {\rm [Fe/H]} \le -0.82$, substantially higher than the cluster mean as can be seen from Fig.~\ref{fig:data}b. The presence of these stars is a fascinating puzzle which we hope to return to in a future contribution. However, since their velocities lie extremely close the cluster mean, it is inconceivable that they are not members, so for the purposes of the current discussion we will treat these stars as dynamical tracers.

The sample whose kinematics we will analyze below consists of all 157 stars marked red in Figure~\ref{fig:vel_radius}, which we will refer to as ``Sample A", and the 9 stars marked in black (Sample B).

\subsection{Rotation}
\label{sec:Rotation}

One of the several interesting results reported by B09 was the signature of rotation in their kinematic sample, which they deduced had a rotational amplitude of $3.26\pm0.85\kms$ directed towards $PA=40\degg9\pm17\degg8$, approximately at right angles to the direction of elongation of the cluster found by B07. The rotation of the cluster is a critical issue, since rotational support can offset pressure support, and obviously influence the kinematic models one has to fit. In reanalyzing the B09 kinematics, we noticed that a significant part of the rotation signal was coming from a small number of stars very close to the cluster center. Because the dispersion increases rapidly towards $R=0$, such stars will generally have velocities quite far from the cluster mean, and if the sample is small (as in the case of B09) stochastic effects can cause a large but spurious rotation signal. To examine this possibility, we analyze sample A, ignoring the stars with $R<0\mcnd5$; we find that the rotation amplitude diminishes to $2.6\kms$ at $PA=-4\deg$ (see Figure~\ref{fig:rotation}). To place confidence limits on the velocity amplitude we undertook a Monte-Carlo experiment, keeping the measured velocities and velocity errors, while setting the position angles of the stars to random values. We found that 5.1\% of the random tests gave fake rotation amplitudes as large as that measured. However, if we limit ourselves to only the DEIMOS sample, the derived rotation amplitude drops to $0.6\kms$ at $PA=-86\degg9$, yet 39\% of random realizations give rise to amplitudes greater than this value. We conclude, therefore, that the B09 rotation measure was largely due to low-number statistics, and that the evidence for significant rotation of the cluster is marginal.

\subsection{Velocity dispersion profile}
\label{sec:sigma_profile}

\begin{figure*}
\begin{center}
\includegraphics[bb= 45 90 560 745, clip, angle=270, width=\hsize]{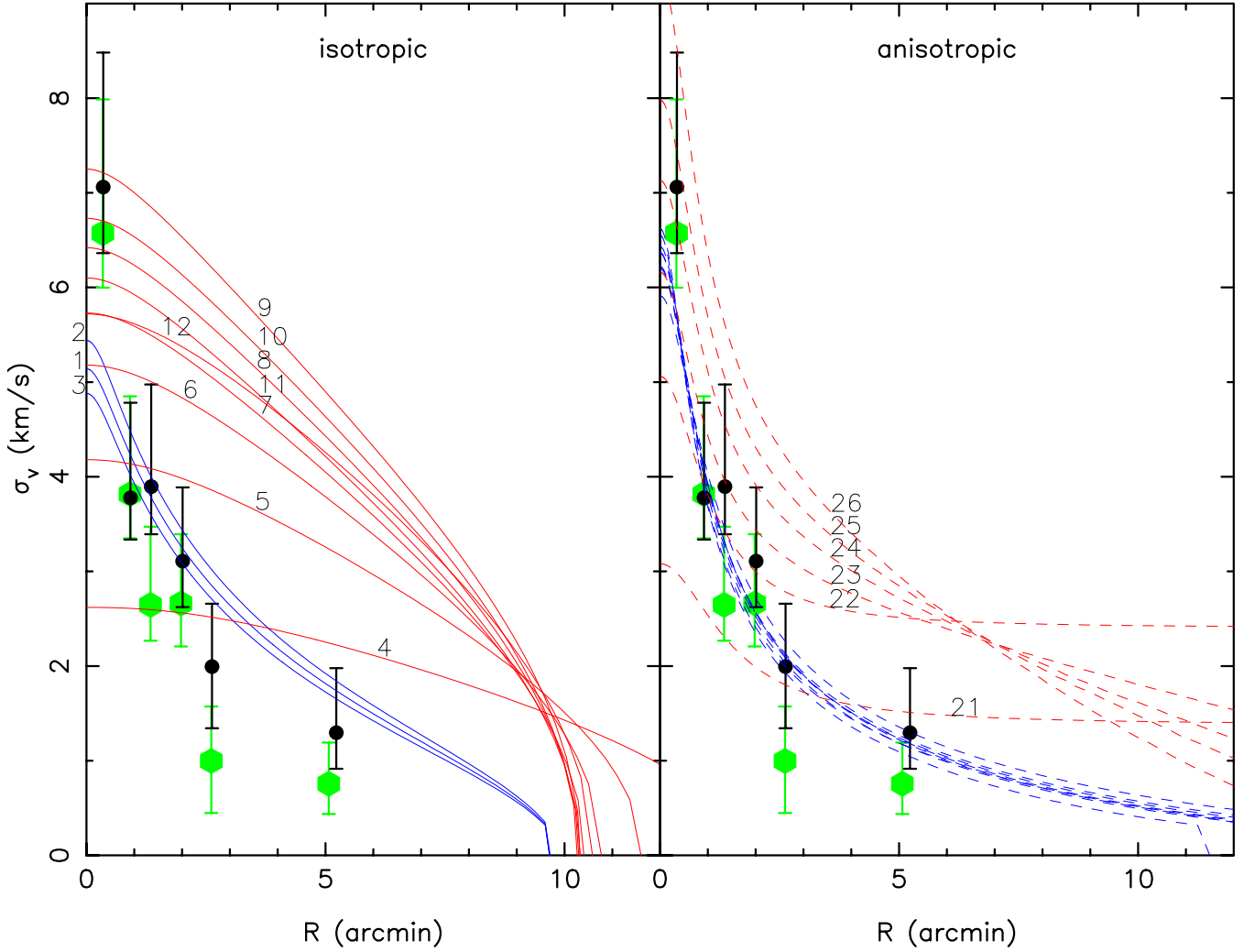}
\end{center}
\caption{The line of sight velocity dispersion profile of the 26 dynamical models is overlaid on the observed profile, derived from sample `A' (green) and sample `A+B' (black). Newtonian models are colored blue, MOND models red, while isotropic models are represented with continuous lines and anisotropic models with dashed lines. As discussed in the text, we expect reality to lie between samples A and A+B; the closest fits are clearly the anisotropic Newtonian models.}
\label{fig:dispersion_profile}
\end{figure*}

While the full likelihood analysis that we present in section~\ref{sec:analysis} is clearly the preferred and most powerful method for ruling out the dynamical models (and their underlying theories of gravity), it may be useful to some readers to visualize the velocity dispersion profile.  To this end we calculate the estimates of the velocity dispersion, using a maximum-likelihood method, under the assumption that the underlying distributions are Gaussian (see Eqn.~3 of \citealt{Pryor:1993p11793}). For sample A, we truncate the Gaussian distributions to account for the effect of the imposed selection function.  The resulting dispersions for samples A and A+B are shown in Fig.~\ref{fig:dispersion_profile} and reported in Table~\ref{tab:dispersions}. We expect reality to lie somewhere between these two estimates.

\begin{table}
\caption{Velocity dispersion estimates for the two samples.}
\label{tab:dispersions}
\hbox{
\begin{tabular}{ccc}
\multicolumn{3}{c}{Sample A} \\ \tableline\tableline
$N$ & $R$ & $\sigma$ \\
       & (arcmin)  & $(\kms)$ \\ 
\tableline \\
 26    &   0.343  &      $6.573^{+1.413}_{-0.575}$\\ \\
 26    &   0.912  &      $3.816^{+1.032}_{-0.467}$\\ \\
 26    &   1.340  &      $2.645^{+0.826}_{-0.378}$\\ \\
 26    &   1.978  &      $2.662^{+0.731}_{-0.454}$\\ \\
 26    &   2.613  &      $0.997^{+0.576}_{-0.549}$\\ \\
 27    &   5.068  &      $0.756^{+0.436}_{-0.318}$\\ \\
\tableline\tableline
\end{tabular}
\begin{tabular}{ccc}
\multicolumn{3}{c}{Sample A+B} \\ \tableline\tableline
$N$ & $R$ & $\sigma$ \\
       & (arcmin)  & $(\kms)$ \\ 
\tableline \\
 27    &   0.352  &      $7.061^{+1.420}_{-0.698}$\\ \\
 27    &   0.920  &      $3.778^{+1.002}_{-0.443}$\\ \\
 27    &   1.355  &      $3.896^{+1.079}_{-0.504}$\\ \\
 27    &   2.011  &      $3.109^{+0.779}_{-0.486}$\\ \\
 27    &   2.625  &      $1.994^{+0.664}_{-0.651}$\\ \\
 31    &   5.226  &      $1.295^{+0.683}_{-0.381}$\\ \\
\tableline\tableline
\end{tabular}
}
\end{table}

\section{Models}
\label{sec:models}

\subsection{Dynamical models}
\label{sec:dynamical_models}

The models we compare to in this paper have been specifically constructed as an extension of those described in \citet[][hereafter SN10]{Sollima:2010p14304}.  The phase-space distribution of stars is given by the distribution function
\begin{eqnarray}
\label{eq_df}
f(E,L)&=&f_{0}~\exp\left(-\frac{L^{2}}{2\sigma_{K}^{2}r_{a}^{2}}\right)\left[\exp\left(-\frac{E}{\sigma_{K}^{2}}\right)-1\right],\nonumber\\ 
f(r,v_{r},v_{t})&=&f_{0}~\exp\left[-\frac{v_{t}^{2}}{2\sigma_{K}^{2}}\left(\frac{r}{r_{a}}\right)^{2}\right]~\times\nonumber\\
& &\left[\exp\left(-\frac{v_{r}^{2}+v_{t}^{2}}{2\sigma_{K}^{2}}-\frac{\psi}{\sigma_{K}^{2}}\right)-1\right]
\end{eqnarray}
\citep{Michie:1963p15300}, where $E$ and $L$ are, respectively, the energy and angular momentum per unit mass, $v_r$ and $ v_t$ are the radial and tangential components of the velocity, the effective potential $\psi$ is the difference between the cluster potential at a given radius $r$ and the potential at the cluster tidal radius $\psi=\phi-\phi_{t}$, $f_{0}$ is a scale factor, and $\sigma_{K}$ is a normalization term which is proportional to the central velocity dispersion.  As usual, the above distribution function has been integrated to obtain the density and the radial and tangential components of the velocity dispersion:
\begin{eqnarray}
\rho(r)&=& 4 \pi \int_{0}^{\sqrt{-2\psi}}\int_{0}^{\sqrt{-2\psi-v_{r}^{2}}}
v_{t}~f(r,v_{r},v_{t}) ~\mathrm{d} v_{t} \mathrm{d} v_{r},\nonumber\\
\sigma_{r}^{2}(r)&=&\frac{4
\pi}{\rho(r)}\int_{0}^{\sqrt{-2\psi}} v_{r}^{2} \int_{0}^{\sqrt{-2\psi-v_{r}^{2}}} v_{t}~ f(r,v_{r},v_{t})
~\mathrm{d} v_{t} \mathrm{d} v_{r},\nonumber\\
\sigma_{t}^{2}(r)&=&\frac{4 \pi}{\rho(r)}\int_{0}^{\sqrt{-2\psi}}
\int_{0}^{\sqrt{-2\psi-v_{r}^{2}}} v_{t}^{3}~f(r,v_{r},v_{t})
~\mathrm{d} v_{t} \mathrm{d} v_{r}.\nonumber
\end{eqnarray}
The above equations can be written in terms of dimensionless quantities by substituting
\begin{eqnarray}
\zeta=\frac{v_{r}^2}{2\sigma_{K}^{2}}, & \eta=\frac{v_{t}^2}{2\sigma_{K}^{2}},\nonumber\\ 
\tilde{\rho}=\frac{\rho}{\rho_{0}}, & \tilde{r}=\frac{r}{r_{c}},\nonumber\\
W=-\frac{\psi}{\sigma_{K}^{2}}, & \ratilde=\frac{r_{a}}{r_{c}},\nonumber
\end{eqnarray}
where $\rho_{0}$ is the central cluster density and
$$r_{c}\equiv\left(\frac{9 \sigma_{K}}{4 \pi G \rho_{0}}\right)^{1/2}$$
is the core radius \citep{King:1966p15212}.  The parameter $\ratilde$ determines the radius at which orbits become more radially biassed.  As $\ratilde \rightarrow\infty$, models become isotropic.  The dimensionless expressions are:
\begin{eqnarray}
\label{eq_models}
\tilde{\rho}(\tilde{r})&=&\frac{\int_{0}^{W}\zeta^{-\frac{1}{2}}\int_{0}^{W-\zeta}e^{-\frac{\eta\tilde{r}^{2}}{\ratilde^{2}}}(e^{W-\eta-\zeta}-1) 
~\mathrm{d}\eta ~\mathrm{d}\zeta}{\int_{0}^{W_{0}}\zeta^{-\frac{1}{2}}\int_{0}^{W_{0}-\zeta}e^{-\frac{\eta\tilde{r}^{2}}{\ratilde^{2}}}(e^{W_{0}-\eta-\zeta}-1) 
~\mathrm{d}\eta ~\mathrm{d}\zeta},\nonumber\\
\sigma_{r}^{2}(\tilde{r})&=&\frac{2\sigma_{K}^{2}\int_{0}^{W}\zeta^{\frac{1}{2}}\int_{0}^{W-\zeta}
e^{-\frac{\eta\tilde{r}^{2}}{\ratilde^{2}}}(e^{W-\eta-\zeta}-1) 
~\mathrm{d}\eta ~\mathrm{d}\zeta}{\int_{0}^{W}\zeta^{-\frac{1}{2}}\int_{0}^{W-\zeta}e^{-\frac{\eta\tilde{r}^{2}}{\ratilde^{2}}}(e^{W-\eta-\zeta}-1) 
~\mathrm{d}\eta ~\mathrm{d}\zeta},\nonumber\\
\sigma_{t}^{2}(\tilde{r})&=&\frac{2\sigma_{K}^{2}\int_{0}^{W}\zeta^{-\frac{1}{2}}\int_{0}^{W-\zeta}
\eta e^{-\frac{\eta\tilde{r}^{2}}{\ratilde^{2}}}(e^{W-\eta-\zeta}-1) 
~\mathrm{d}\eta ~\mathrm{d}\zeta}{\int_{0}^{W}\zeta^{-\frac{1}{2}}\int_{0}^{W-\zeta}e^{-\frac{\eta\tilde{r}^{2}}{\ratilde^{2}}}(e^{W-\eta-\zeta}-1) 
~\mathrm{d}\eta ~\mathrm{d}\zeta}.\nonumber\\
\end{eqnarray}
The above equations allow one to derive the density and the velocity dispersions once the potential profile $W(\tilde{r})$ is known.  This last quantity is linked to the density by the Poisson equation (or its MOND modification; Eqn.~\ref{eq_mond}) with the boundary conditions at the centre
\begin{eqnarray}
\label{bound_eq}
W&=& W_{0} \, , \nonumber\\
\frac{\mathrm{d} W}{\mathrm{d}\tilde{r}}&=&0 \, .\nonumber
\end{eqnarray}
Here (and below, when not specified otherwise) we adopt the ``simple" interpolating function $\mu(x)=x/(1+x)$ \citep{Famaey:2005p15435}.  In spherical symmetry Equation~(\ref{eq_mond}) can be written in the form
$$ \frac{1}{r^{2}}\frac{\mathrm{d}}{\mathrm{d} r}\left[r^{2}\mu\left({{1\over a_{0}}\left|{\mathrm{d} \psi\over \mathrm{d} r}\right|}\right)\frac{\mathrm{d}\psi}{\mathrm{d} r}\right]=4\pi G
\rho \, ,$$
or, using dimensionless quantities,
\begin{equation}
\label{poiss_eq}
\frac{1}{\tilde{r}^{2}}\frac{\mathrm{d}}{\mathrm{d}\tilde{r}}\left[\tilde{r}^{2}
\mu\left({\chi\left|{\mathrm{d} W \over \mathrm{d} \tilde{r}}\right|}\right)
\frac{\mathrm{d} W}{\mathrm{d}\tilde{r}}\right]= -9\tilde{\rho},
\end{equation}
where $\chi\equiv\sigma_{K}^{2} / a_{0}r_{c}$ is a dimensionless parameter, which is smaller for systems closer to the deep-MOND regime (see SN10).  For a given choice of ($W_{0},\chi, \ratilde$), Equations \ref{eq_models} and \ref{poiss_eq} have been integrated to obtain the 3D density and the radial and tangential components of the velocity dispersion.  As a last step, the above profiles have been projected on the plane of the sky to obtain the surface mass density
\begin{equation}
\Sigma_*(R)=2\int_{R}^{\tilde{r_{t}}}
\frac{\tilde{\rho}\tilde{r} \mathrm{d} \tilde{r}}{\sqrt{\tilde{r}^{2}-R^{2}}}\nonumber
\end{equation}
and the line-of-sight velocity dispersion
\begin{equation}
\sigma_{v}^{2}(R)=\frac{1}{\Sigma_*(R)}\int_{R}^{\tilde{r_{t}}}
\frac{\tilde{\rho}\left[2\sigma_{r}^{2}\left(\tilde{r}^{2}-R^{2}\right)+\sigma_{t}^{2}R^{2}\right] \mathrm{d} \tilde{r}}{\tilde{r}\sqrt{\tilde{r}^{2}-R^{2}}} \, .\nonumber
\end{equation}

The described procedure is straightforward and produces spherically symmetric self-consistent equilibrium models, truncated at a tidal radius $\tilde{r}_{t}$. Note that spherical symmetry is justified as the evidence to the contrary (a measurement of $\epsilon =0.19\pm0.15$ or $\epsilon =0.14\pm0.07$ by B07 --- depending on the sample choice) is not very strong. Also, the above models assume a single distribution function for all stars, regardless of their masses. This can be considered a good approximation since mass segregation effects are expected to be negligible in NGC 2419. This can be deduced by the fact that the relaxation time is significantly larger than the cluster age ($t_{rh}/t_{age}\sim$3.5; \citealt{Harris:1996p11754, MarinFranch:2009p15556}) and it is confirmed by the lack of radial segregation of the BSS population (D08).

As will be described in Sect. \ref{sec:analysis}, the full likelihood analysis performed here requires for each model $j$ the knowledge of the distribution of projected velocities (instead of simply its second-order moment). This function, hereafter referred to as $f^j(v,R)$, can be calculated as
\begin{eqnarray}
&&f^j(v,R)=\int_{0}^{r_{t}}\frac{\rho(r) r}{\sqrt{r^{2}-R^{2}}}
\int_{0}^{\sqrt{-2\psi-v^2}}\nonumber\\
&&\int_{0}^{\sqrt{-2\psi-v^2-v_{x}^2}} f(r,v_{r}',v_{t}')~\mathrm{d}v_{y} \mathrm{d}v_{x} \mathrm{d}r \, ,
\end{eqnarray}
adopting the following change of variables
\begin{eqnarray}
v_{r}'&=&v~\sin\alpha +v_{x}~\cos\alpha,\nonumber\\
v_{t}'&=&[(v~\cos\alpha -v_{x}~\sin\alpha)^{2}+v_{y}^{2}]^{1/2} \, ,\nonumber
\end{eqnarray}
where
$$\alpha=\arccos(R/r) \, .$$

\subsection{Stability analysis}
\label{sec:stability}

It is well known that extremely radially anisotropic Newtonian models are prone to bar instability \citep[e.g.,][]{Fridman:1984p17072}. \citet{Nipoti:2011p15570} have recently shown that this radial-orbit instability operates also in MOND, thus limiting the fraction of radial orbits in realistic models. To test the stability of our models we run a set of N-body simulations using the MOND $N$-body code \NMODY\ \citep{Nipoti:2007p15571,Londrillo:2009p15575}, which can be used to follow the evolution of either MOND or Newtonian collisionless stellar systems. The use of a collisionless N-body code is justified by the fact that NGC~2419 has a long relaxation time (see \S\ref{sec:dynamical_models},~\S\ref{sec:binaries} and B08). The stability analysis was performed following \citet{Nipoti:2011p15570}, and we refer the reader to that paper for details. The initial conditions of the simulations consist of an N-body system of $8\times10^5$ particles, which are distributed in phase space with a standard rejection technique using the analytic distribution function given in Equation~(\ref{eq_df}).  We verified that the isotropic or mildly anisotropic (i.e. with large anisotropy radius $r_a$ as compared to the core radius $r_c$) MOND and Newtonian N-body models are stable. Then we explored the behavior of the models for decreasing $r_a$, finding in each case the minimum anisotropy radius for stability $r_{as}$. For Newtonian models we found $r_{as}/r_c\sim 0.9$. For MOND models the value of $r_{as}/r_c$ can depend on the mass-to-light ratio $(M/L)_V$, so for each value of $(M/L)_V$ we ran a set of MOND simulations with decreasing $r_{a}/r_c$, thus determining $r_{as}/r_c$. The most anisotropic stable MOND Michie models, for the considered $(M/L)_V$ are those reported in Table~\ref{tab:models}, with $r_{as}/r_c$ in the range $1.4-1.6$.  In terms of the half-mass radius $r_{half}$, we have $r_{as}/r_{half}\sim0.5$ in the Newtonian case and $r_{as}/r_{half}\sim0.8-1$ in MOND: thus Newtonian models of NGC~2419 can have significantly smaller $r_{as}$ than corresponding MOND models, in agreement with the findings of \citet{Nipoti:2011p15570} for a different class of stellar systems. It is also interesting to consider the Fridman-Polyachenko-Shukhman parameter $\xi \equiv 2 T_r/T_t$ \citep[where $T_r$ and $T_t$ are the radial and tangential components of the kinetic energy tensor; see][]{Fridman:1984p17072} and of the quantity $\xi_{half}$, defined as twice the ratio of radial to tangential kinetic energy within $r_{half}$ (so $\xi=\xi_{half}=1$ for isotropic systems). The maximum values for stability are $\xi_{s}\sim1.9$ and $\xi_{half,s}\sim1.5$ in Newtonian gravity, and $\xi_{s}\sim2- 2.1$ and $\xi_{half,s}\sim 1.4-1.5$ in MOND.

\subsection{Binaries}
\label{sec:binaries}

\begin{figure*}
\begin{center}
\includegraphics[angle=0, bb= 45 90 560 745, clip, angle=270, width=\hsize]{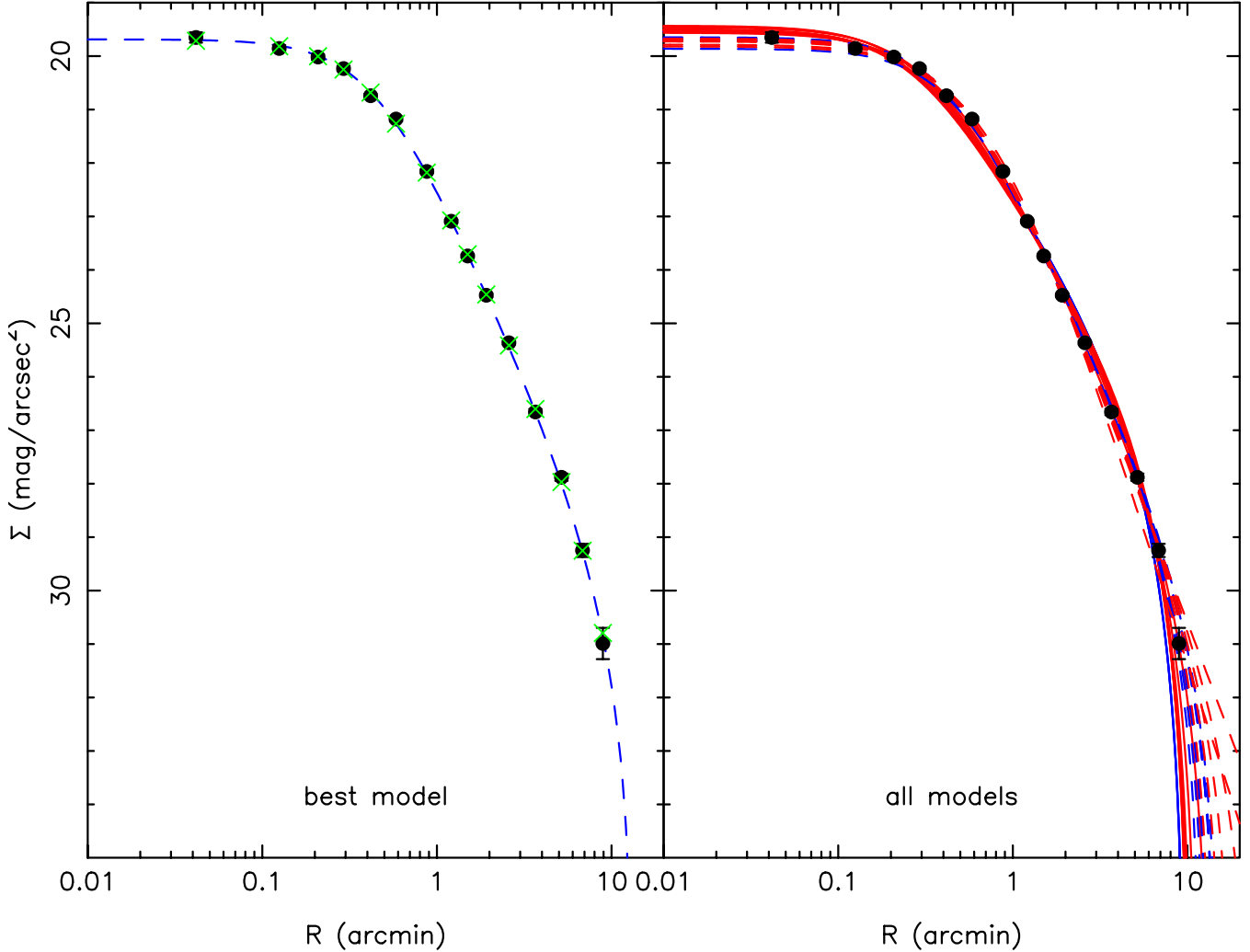}
\end{center}
\caption{The star counts profile derived in \S\ref{sec:Observations} (and listed in Table~\ref{tab:starcounts}) is marked by black dots with their associated uncertainties. The blue dashed line on the left-hand panel shows the best-fit model (the anisotropic Newtonian model \#17). The green crosses mark the values of this model integrated over the same radial bin as each observed data point, taking into account the complex spatial selection function (gaps between CCDs, etc). The right-hand panel displays the full library of Michie models; the colors and line styles have the same meaning as in Fig.~\ref{fig:dispersion_profile}.}
\label{fig:SB_models}
\end{figure*}

\begin{figure*}
\begin{center}
\includegraphics[angle=0, bb= 45 90 560 745, clip, angle=270, width=\hsize]{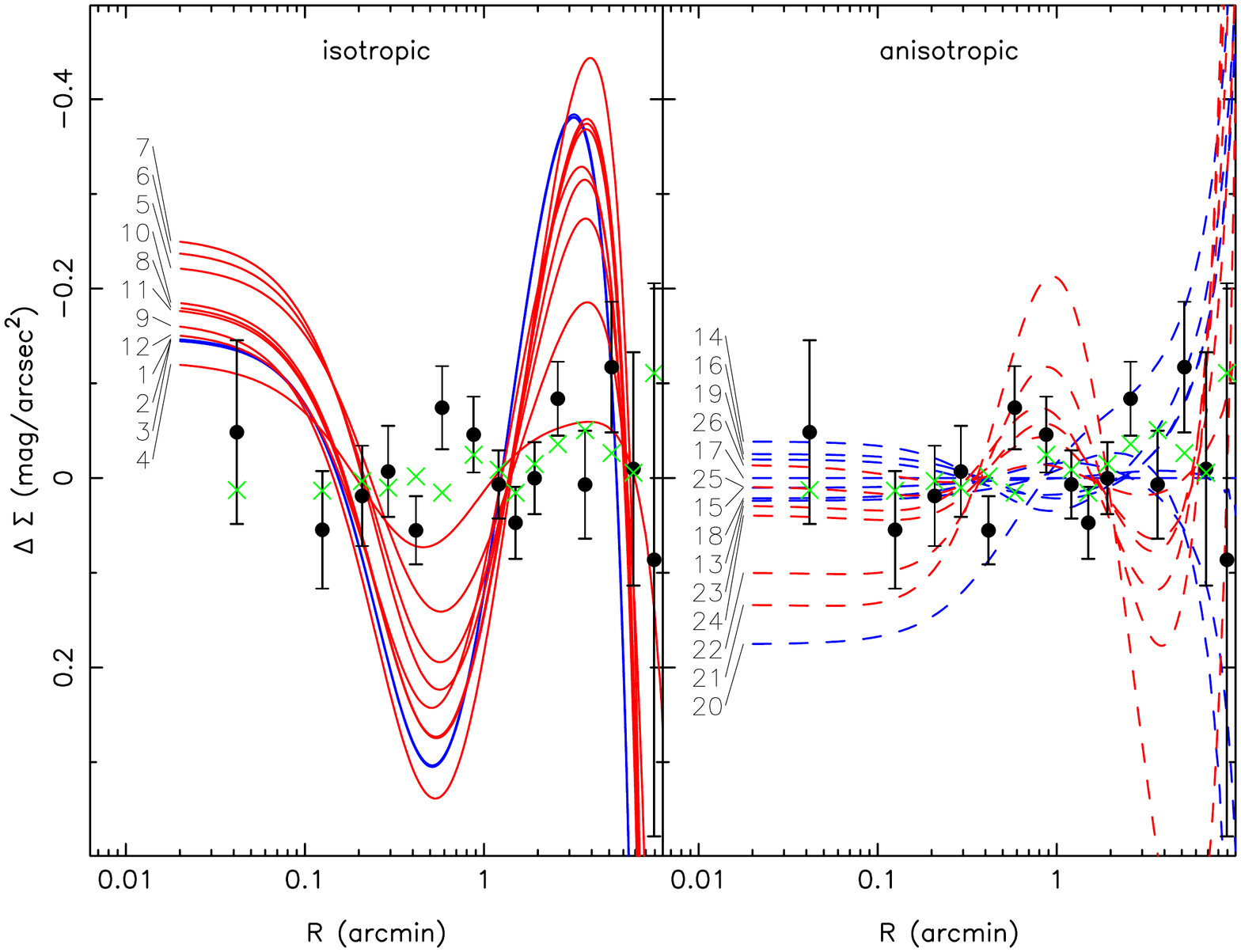}
\end{center}
\caption{The surface brightness residuals from the model (\#17) that fits best the surface brightness profile. The identification numbers of the models are indicated. The meaning of the dots, crosses, colors and line styles are the same as in Fig.~\ref{fig:SB_models}.}
\label{fig:SB_residuals}
\end{figure*}

Before proceeding to the detailed comparison of the above models to the new data presented in this contribution, we first consider the likely effect of binary stars in the sample on the distribution of observed kinematics.  Following the seminal paper by \citet[][hereafter RB97]{Rubenstein:1997p15036}, the most comprehensive recent attempts to determine the binary fraction ($b_f$) in globular clusters have been performed by modeling the spread of the Main Sequence in very accurate Hubble Space Telescope color-magnitude diagrams, taking into account the observational effects (in particular blending) by means of extensive artificial star experiments \citep{Bellazzini:2002p15037, Sollima:2007p15042, Milone:2008p17022}. These results are clearly dependent on the assumed distribution of mass ratios $q$; however, essentially all studies try to overcome this problem by using several different distributions.

From these analyses globular clusters have been found to contain copious quantities of binaries. RB97 found $15\% \le b_f \le 38\%$ in the core of NGC~6752, which is a {\it much} denser cluster than NGC~2419 (with logarithm of central luminosity density $\log \rho_0 =4.9 \lsun/{\rm pc}^3$ as opposed to $1.5 \lsun/{\rm pc}^3$), and $b_f \le 16\%$ outside. Similarly, \citet{Bellazzini:2002p15037} found $10\% \le b_f \le 20\%$ in the core of NGC~288 (which with $\log \rho_0 = 1.8 \lsun/{\rm pc}^3$ is closer in density to NGC~2419) and $\le 10\%$ outside the core (consistent with 0\%). With a larger sample of 13 low density clusters, \citet{Sollima:2007p15042} concluded that the {\it minimum} $b_f$ in the central areas they sampled was $\ge 6\%$ in {\it all} the considered clusters, but may be as large as 20\% in some cases. Most of their best fit values range between 10\% and 20\%, with peaks of 50\% (Terzan 7). The cluster with the central density most similar to NGC~2419 in their sample (NGC~6101) has a minimum $b_f =9\%$ and best fit values ranging from 15\% to 21\%, depending on the assumed distribution of $q$.

We know that binaries must also be present in NGC2419, since the cluster has numerous Blue Stragglers Stars (BSS) \citep[][hereafter D08]{Dalessandro:2008p15030} while collisions are nearly impossible (because of the long relaxation time), thus essentially all BSS must come from the evolution of Primordial Binaries. As a comparison, the fraction of BSS normalized to the total luminosity is double that in $\omega$-Cen (whose binary fraction has been estimated to be 13\%, \citealt{Sollima:2007p15042}). On the other hand, the ratio of BSS to Horizontal Branch stars $N_{BSS}/N_{HB} =0.3$. The same parameter is 0.07 in M13, 0.3 in M3 and 0.92 in NGC~288 (\citealt{Ferraro:2003p15038}, computed in a homogeneous way). As already reported in Sect. \ref{sec:models}, D08 found that the BSS show no sign of radial segregation, supporting the idea that two-body relaxation is highly inefficient in NGC~2419. This is the observational foundation for two key assumptions of our analysis: first, this implies that $b_f$ is constant with radius, making it easier to model a population of binaries; second, and more importantly, it implies that the mass to light ratio of the cluster is constant with radius. If mass segregation had been at work, as in most other clusters, a larger fraction of low-mass stars would have been present in the outer regions, enhancing $(M/L)_V$ there with respect to the core.

\citet{Sollima:2008p15044} found that the fraction $F$ of BSS to main-sequence stars (computed in a way that may be difficult to apply to NGC~2419) correlates very well with $b_f$, and $M_V$, in the sense that brighter clusters have a smaller $F$, and, as a consequence, $b_f$.  So, while there is a possibility that $b_f$ may be lower than average in NGC~2419, a fraction $b_f$ as large as 20$\%$ is clearly possible and cannot be discounted.

Note that the effect of binaries on the velocity dispersion should be larger in the outer parts of the system, enhancing the velocity dispersion as would be expected if MOND were correct.

\subsection{Modeling the binary fraction}
\label{sec:binaries_model}

To construct a model for the effect of the binary population on the observed kinematics, we run a Monte Carlo simulation. Following McConnachie \& Cote (2010) the projected motion of the primary component in a binary system is given by
$$v=\frac{2\pi a_{1} \sin i}{P(1-e^{2})^{1/2}}[\cos(\theta+\omega)+e \cos
\omega]$$ where $a_{1}$ 
is the semi-major axis of the primary component, $P$ is the orbital period, $e$ the eccentricity, $i$ the inclination angle to the line-of-sight, $\theta$ the phase from the periastron and $\omega$ the longitude of the periastron. The semi-major axis has been calculated using the third Kepler law
$$a_{1}=\frac{1}{1+\frac{m_{1}}{m_{2}}}\left[\frac{P^{2}G(m_{1}+m_{2})}{4\pi^{2}}\right]^{1/3}$$
where $m_{1}$ and $m_{2}$ are the masses of the primary and secondary component.  For each simulated binary we exctracted randomly a combination of ($m_{1},m_{2}, P,e,\theta,\omega,i$) from suitable distributions and derived the corresponding projected velocity $v$.  We adopted a fixed mass for the primary component of $m_{1}=0.83~M_{\odot}$ (i.e. the typical mass of a RGB star, calculated by comparing the color-magnitude diagram of NGC~2419 with a suitable isochrone of Marigo et al. 2008) and extracted the mass of the secondary component from the mass-ratio distribution by \citet{Fisher:2005p15039}. We followed the prescriptions of \citet{Duquennoy:1991p15035} for the distribution of periods and eccentricities. We removed all those binaries whose corresponding semi-axis lie outside the range $a_{min}<a<100$~AU where $a_{min}$ is linked to the radius of the secondary component (according to \citealt{Nelson:1986p15034}). The distribution of the angles ($i,\theta,\omega$) has been chosen according to the corresponding probability distributions (${\rm Prob}(i)\propto \sin~i;~{\rm Prob}(\theta)\propto \dot{\theta}^{-1};~{\rm Prob}(\omega)={\rm constant}$).
 
\begin{figure}
\begin{center}
\includegraphics[bb= 50 90 560 740, angle=270, clip, width=\hsize]{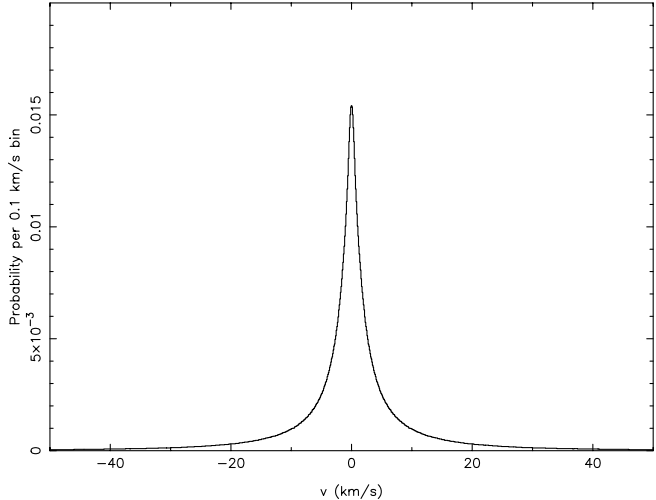}
\end{center}
\caption{The velocity distribution model for binaries. To account for a possible binary fraction $b_f$, this distribution is convolved with the line of sight distribution of the dynamical models as detailed in Eqn.~\ref{eqn:binaries}.}
\label{fig:binaries}
\end{figure}

The resulting distribution $g(v)$ is shown in Figure~\ref{fig:binaries}. The velocity distribution corrected for binaries $f^j_b(v,R_i)$ is derived by convolving $g(v)$ with the dynamical models. Since $b_f$ is expected not to change significantly with radius in NGC~2419, as discussed above, we simply take:
\begin{equation}
f^j_b(v,R_i)=(1-b_f) f^j(v,R_i)+b_f \int{f^j(v',R_i) g(v'-v) \mathrm{d}v'} \, ,
\label{eqn:binaries}
\end{equation}
where $R_i$ is the projected radius of star $i$ in the sample.  Note that although the wings of the distribution in Figure~\ref{fig:binaries} extend to large velocities, the central peak at $v=0$ is very prominent, so the effect of convolving this distribution with the dynamical models should be small if the binary fraction is small.

\begin{figure}
\begin{center}
\includegraphics[bb= 50 90 560 740, angle=270, clip, width=\hsize]{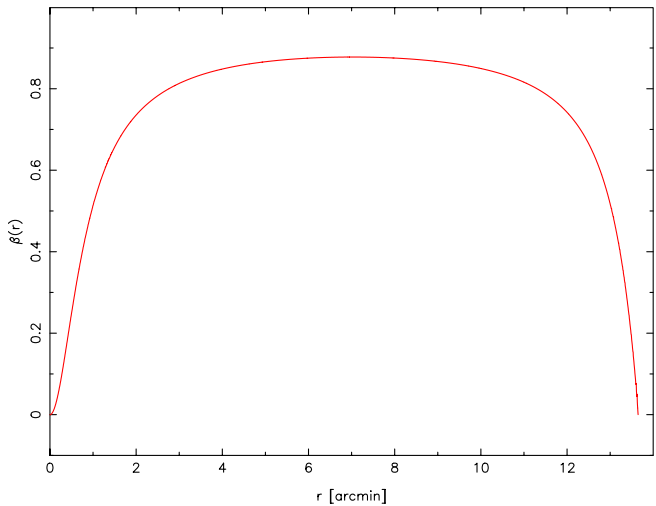}
\end{center}
\caption{The anisotropy profile for model \#17.}
\label{fig:betamodel17}
\end{figure}

\section{Analysis}
\label{sec:analysis}

The final step in constructing the line of sight velocity distribution models is to convolve $f^j_b(v,R_i)$ by a Gaussian function of dispersion equal to the estimated velocity uncertainty $\delta v_i$ of star $i$. This results in the predicted probability distribution $f^j_p(v,R_i)$. The likelihood of each dynamical model $j$ can now finally be calculated via:
\begin{equation}
\ln L^j = \sum_{i=1}^{n} \ln[f^j_p(v_i,R_i)] - \sum {{(\Sigma^j_p - \Sigma)^2}\over{2 \delta\Sigma^2}}   \, ,
\label{eqn:likelihood}
\end{equation}
where we have substituted the actual observed radial velocity measurements $v_i$ into the probability distribution. The second sum in Equation~\ref{eqn:likelihood} is performed over the surface brightness measurements $\Sigma$ for which the model predictions $\Sigma^j_p$ of each model $j$ were displayed above in Fig.~\ref{fig:SB_models}. In calculating the likelihood of the surface brightness predictions we obviously are assuming a Gaussian model with dispersion equal to the estimated measurement uncertainties $\delta\Sigma$. Note that the radial positions of the spectroscopic stars are not drawn in a fair manner, so their positional information cannot be used in the likelihood calculation. The first term on the right hand side of Eqn.~\ref{eqn:likelihood} answers the question ``given the velocity measurements at the radii of the kinematic data, what is the likelihood that the data were drawn from the same distribution as the model?''. If instead we had adopted a $\chi^2$ fit to the velocity dispersion measures, we would effectively have been assuming that the velocity data sample the radial bins in a fair way. That is not a good assumption.

To assess the effect of contamination in the stellar sample, we perform the likelihood analysis on the full data set (samples A+B), but also on the more restrictive sample A. For this latter case, we implement a window function in the likelihood calculation, curtailing the wings of the model velocity distributions $f^j_p$ to within the selection region of the sample (see Figure \ref{fig:vel_radius}). The model distributions are re-normalised to compensate for the discarded velocity parameter space. Table~\ref{tab:models} lists the model number $j$, the adopted theory of gravity, the interpolation function $\mu$ (for MOND), the assumed Heliocentric distance $d$, and the structural parameters of the cluster model. These parameters are the anisotropy radius $r_a$, the core radius $r_c$, the half-mass radius $r_h$, the tidal radius $r_t$, and the central line of sight velocity dispersion $\sigma_0$. The (base-10) logarithm of the ratio of likelihoods of model $j$ to the best model is listed in the columns marked $\log{{L_{max}}\over{L}}$, both for samples A and A+B. We also list the corresponding $(M/L)_V$ (in Solar units); for the Newtonian case, the quoted values are the best-fit results, while for MOND they correspond to the input (fixed) value that was simulated. The derived binary fraction $b_f$ is also given.

Inspection of Table~\ref{tab:models} shows \#17 to be the best representation of the cluster among these Michie models.  The preferred $(M/L)_V$ transpires to be identical to that derived by \citet{McLaughlin:2005p15121}.  To investigate whether this best fitting dynamical model is an {\it acceptable} representation, we undertook a Monte-Carlo test, drawing artificial stars from the model and computing the sample likelihood. Each of the 10000 artificial samples is required to have the same number of stars as the actual kinematic sample, and the artificial stars are selected at the model radii of the observed stars. We find that 61\% of these randomly-drawn samples have likelihood larger than that measured for model \#17 from the observed sample. This indicates that the best-fitting Michie model is an excellent representation of the cluster\footnote{Had we found a very high fraction (say 90\%) of random realizations with higher likelihood than that measured, it would have cast doubt on the model, while a very low fraction (say 10\%) would have called the reliability of the uncertainty estimates into question.}.

\begin{table*}
\begin{center}
\caption{The cluster structural parameters, and the adopted theory of gravity are listed for the 26 Michie models. The final six columns show the corresponding best fits given the data in sample A and sample A+B. The likelihoods are listed as ratios with respect to the best-fitting model (for each sample separately). The best-fit binary fractions are also given. Thus for sample A, model \#1 is $10^{129.9}$ less likely than model \#17. For the Newtonian models the $(M/L)_V$ listed are best-fit values, whereas in MOND they correspond to an input of the simulation.}
\label{tab:models}
\begin{tabular}{|ccccrrrrc|ccc|ccc|}
\tableline\tableline
$j$           & Gravity  & $x/\mu(x)$         &$d$\,\,& $r_a/r_c$      & $r_c$ \, & $r_h$ \,&$r_t$\,\,& $\sigma_0$ & $\log{{L_{max}}\over{L}}$ & $\big({{M}\over{L}}\big)$ & $(b_f)$ & $\log{{L_{max}}\over{L}}$ & $\big({{M}\over{L}}\big)$ & $(b_f)$ \\
              &          &                    &$(\kpc)$&               & $(\pc)$  & $(\pc)$ &$(\pc)$\,&$({\rm {{km}\over{s}}})$ & & & & & &   \\
              &          &                    &       &                &          &         &         &          &\multicolumn{3}{c|}{Results sample A} & \multicolumn{3}{c|}{Results sample A+B}\\
\tableline
1             & Newton   &    ---             & 87.5  &  $\infty$      &     6.88 &   28.62 &  245.80 &    5.14  & 129.9 & 1.829 & 0\% &  129.1& 1.694 & 8\% \\
2             & Newton   &    ---             & 96.25 &  $\infty$      &     7.57 &   31.49 &  270.46 &    5.44  & 141.6 & 1.632 & 0\% &  140.8& 1.517 & 8\% \\
3             & Newton   &    ---             & 78.75 &  $\infty$      &     6.20 &   25.79 &  221.51 &    4.88  & 119.0 & 2.023 & 0\% &  118.2& 1.903 & 8\% \\
4             & MOND     &    $1+x$           & 87.5  &  $\infty$      &    15.42 &   24.67 &  350.01 &    2.62  &  21.4 & 0.100 & 0\% &   11.4& 0.100 &28\% \\
5             & MOND     &    $1+x$           & 87.5  &  $\infty$      &    10.03 &   24.97 &  294.65 &    4.18  &  31.6 & 0.500 & 0\% &   31.1& 0.500 & 5\% \\
6             & MOND     &    $1+x$           & 87.5  &  $\infty$      &     8.65 &   25.86 &  273.54 &    5.18  &  47.7 & 1.000 & 0\% &   48.6& 1.000 & 0\% \\
7             & MOND     &    $1+x$           & 87.5  &  $\infty$      &     8.09 &   26.28 &  264.21 &    5.73  &  64.0 & 1.346 & 0\% &   66.1& 1.346 & 0\% \\
8             & MOND     &    $1+x$           & 87.5  &  $\infty$      &     7.86 &   27.60 &  261.47 &    6.42  & 232.2 & 1.903 & 0\% &  236.3& 1.903 & 0\% \\
9             & MOND     &    $1+x$           & 87.5  &  $\infty$      &     7.37 &   28.88 &  262.10 &    7.25  & 887.7 & 2.691 & 0\% &  894.8& 2.691 & 0\% \\
10            & MOND     &    $1+x$           & 96.25 &  $\infty$      &     8.62 &   30.26 &  286.75 &    6.73  & 268.8 & 1.903 & 0\% &  274.1& 1.903 & 0\% \\
11            & MOND     &    $1+x$           & 78.75 &  $\infty$      &     7.09 &   24.88 &  235.86 &    6.10  & 195.0 & 1.903 & 0\% &  198.0& 1.903 & 0\% \\
12            & MOND     &    $\sqrt{1+x^2}$  & 87.5  &  $\infty$      &     7.45 &   26.46 &  221.90 &    5.80  &  66.3 & 1.903 & 0\% &   68.8& 1.903 & 0\% \\
13            & Newton   &    ---             & 87.5  &  0.9           &    13.70 &   24.11 &  292.23 &    6.62  &   9.3 & 2.109 & 0\% &    9.2& 2.065 & 8\% \\
14            & Newton   &    ---             & 87.5  &  1.1           &    12.31 &   23.88 &  359.96 &    6.55  &   1.7 & 2.065 & 0\% &    1.7& 2.023 & 8\% \\
15            & Newton   &    ---             & 87.5  &  1.3           &    12.20 &   23.91 &  327.61 &    6.37  &   8.2 & 2.023 & 0\% &    8.0& 1.981 & 8\% \\
16            & Newton   &    ---             & 87.5  &  1.4           &    11.66 &   23.79 &  361.16 &    6.43  &   1.0 & 1.942 & 0\% &    1.0& 1.903 & 8\% \\
17            & Newton   &    ---             & 87.5  &  1.5           &    11.70 &   23.87 &  331.27 &    6.36  &   0.0 & 1.903 & 0\% &    0.0& 1.903 & 8\% \\
18            & Newton   &    ---             & 87.5  &  1.6           &    11.70 &   23.87 &  314.19 &    6.22  &   9.1 & 1.942 & 0\% &    8.9& 1.903 & 8\% \\
19            & Newton   &    ---             & 87.5  &  1.7           &    11.13 &   23.93 &  364.33 &    6.20  &   1.1 & 1.942 & 0\% &    1.0& 1.903 & 8\% \\
20            & Newton   &    ---             & 87.5  &  2.0           &    11.67 &   26.02 &  378.50 &    5.91  &  12.3 & 1.903 & 0\% &   12.1& 1.866 & 7\% \\
21            & MOND     &    $1+x$           & 87.5  &  1.4           &    23.33 &   24.50 &  23330. &    3.08  &  59.8 & 0.100 & 0\% &   53.9& 0.100 &28\% \\
22            & MOND     &    $1+x$           & 87.5  &  1.4           &    17.05 &   24.55 &  17050. &    5.06  &  14.2 & 0.500 & 0\% &   13.3& 0.500 & 4\% \\
23            & MOND     &    $1+x$           & 87.5  &  1.4           &    14.95 &   23.92 &  587.00 &    6.16  &   7.5 & 0.861 & 0\% &    6.7& 0.861 & 0\% \\
24            & MOND     &    $1+x$           & 87.5  &  1.5           &    13.95 &   23.99 &  460.87 &    7.13  &   5.7 & 1.346 & 0\% &    5.3& 1.346 & 0\% \\
25            & MOND     &    $1+x$           & 87.5  &  1.6           &    13.02 &   23.83 &  403.28 &    7.98  &   5.3 & 1.903 & 0\% &    5.8& 1.903 & 0\% \\
26            & MOND     &    $1+x$           & 87.5  &  1.5           &    12.48 &   23.84 &  346.91 &    9.30  &   4.6 & 2.691 & 0\% &    6.7& 2.691 & 0\% \\
\tableline\tableline
\end{tabular}
\end{center}
\end{table*}

\section{Discussion}
\label{sec:discussion}

The analysis presented above shows clearly that the massive halo globular cluster NGC~2419, which is sufficiently distant not to be significantly affected by the external field of the Milky Way, and is sufficiently extended so that a substantial fraction of its member stars experience extremely low accelerations, conforms to our expectations of a plausible dynamical system in Newtonian gravity. The Michie model \#17 with $(M/L)_V=1.903$ and $r_a=1.5 r_c$, which has the anisotropy profile shown in Fig.~\ref{fig:betamodel17}, is found to give an excellent representation of the cluster. Isotropic Michie (i.e. King) models in Newtonian gravity are a factor of $10^{11}$ less likely than the best anisotropic model, and can be firmly excluded. The isotropic King models in MOND for any plausible mass to light ratio are also completely excluded, as their likelihoods are exceedingly low compared to the best anisotropic Newtonian model. The anisotropic Michie models in MOND fare better: for sample A, the most likely such model is \#26, but this is still a factor of 40000 less likely than model \#17, while for the full sample A+B, the best MOND model is \#24, but again it is much less likely (by a factor of $2\times 10^5$) than model \#17. Therefore we also reject the anisotropic Michie models in MOND.

Stellar population models by \citet{Maraston:1998p17067}, \citet{Percival:2009p17086} and \citet{Buzzoni:1989p17070}, with age and metallicity similar to NGC~2419, predict a mass to light ratio in the range $(M/L)_V = 1.6$ -- 3.0, dependent on the assumed IMF and to a lesser extent, on the particular evolutionary synthesis model. Values of $(M/L)_V \sim 3$ result from a Salpeter IMF, while values of $(M/L)_V \simlt 2$ arise from assuming a Kroupa IMF. Our best-fit model \#17 has a mass to light ratio that seems perfectly plausible in this context.

Sample A was constructed to avoid velocity outliers, so it is natural that the MCMC algorithm returned $b_f=0$ for that set. For the full sample A+B, the most likely Newtonian models have $b_f \sim 8$\%, in reasonable agreement with the findings of \citet{Sollima:2007p15042}. The best MOND solutions however, favor $b_f=0$; clearly higher $b_f$ would increase the effective velocity dispersion, producing a more acute discrepancy with the kinematic data. It is interesting to note that including a plausible binary fraction makes the situation significantly worse for MOND.

\begin{figure}
\begin{center}
\includegraphics[bb= 50 90 560 740, angle=270, clip, width=\hsize]{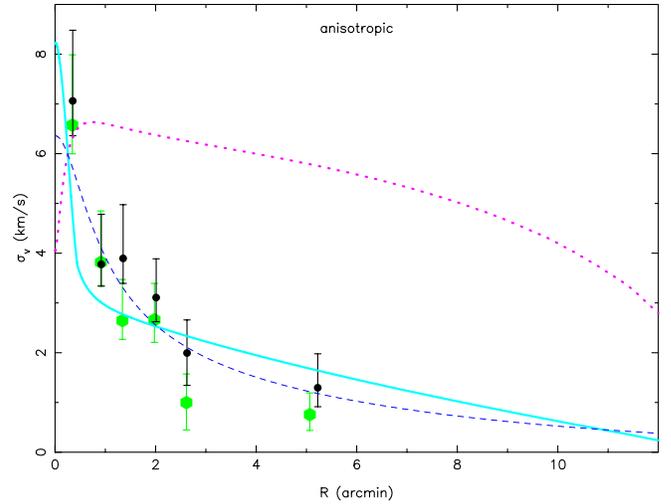}
\end{center}
\caption{The line of sight velocity dispersion profile of two extreme kinematic toy models of the cluster in MOND are compared to the dynamical Newtonian Michie model \#17 (dashed blue line). A purely tangential model (magenta dotted line) is inconsistent with the data beyond the central regions, but a maximally radial model (full turquoise line) appears more promising. The data are represented with the same coding as in Fig.~\ref{fig:dispersion_profile}.}
\label{fig:toymodels}
\end{figure}

The analysis above has considered a realistic family of dynamical models --- the Michie models --- which are consistent in the sense that they have a positive distribution function (see Section~\ref{sec:dynamical_models}).  Michie models are generic models that fit real systems in Newtonian gravity (including this cluster), and reflect the expectation that the outer regions retain some memory of their initial collapse with higher radial than tangential motions. Alternatively, one mechanism that could give rise to such radial anisotropy, is if the cluster expands violently via gas expulsion shortly after its formation \citep{Baumgardt:2008p17865}. The considered models were also tested for stability, as detailed above. Although the Michie models are fairly general, the cluster could of course have some other anisotropy profile, but the difficulty is finding an appropriate self-consistent dynamical model that fits the system.

It is perhaps instructive first to see what anisotropy profile would help bring the line of sight velocity dispersion profile closer to the data, without worrying about the dynamical consequences. Figure~\ref{fig:toymodels} compares the binned data previously displayed in Fig.~\ref{fig:dispersion_profile}, with two toy models. Here we have taken the density profile of the Michie model \#17 as a given, and assume a constant mass to light ratio, as supported by the results of D08. The magenta dotted line displays a model in MOND (with $(M/L)_V=1.9$) which has perfectly tangential orbits. While this approximates the value of the central line of sight velocity dispersion, it vastly over-predicts that at larger radius. In contrast, a highly radially anisotropic model in MOND, that respects the Global Density Slope-Anisotropy Inequality \citep[GDSAI,][]{Ciotti:2010p15374},
\begin{equation}
2\beta(r) \le -{{\mathrm{d} \ln \rho(r)}\over{\mathrm{d} \ln(r)}} \,,
\label{eqn:GDSAI}
\end{equation}
which is a requirement for consistency of a wide range of models, is shown with the full turquoise line (for the particular choice of $(M/L)_V=0.5$). Here, $\rho$ is the density and $\beta \equiv 1- \overline{v_\theta^2}/\overline{v_r^2}$ is the anisotropy parameter. Clearly, the choice of low $(M/L)_V$ and high radial anisotropy is more promising. However, we recall that the GDSAI is a necessary (but not sufficient) condition for consistency, so we have no guarantee that there is a consistent model with such an extreme anisotropy profile, not to mention the issue of stability and the astrophysical inconsistency of such a low $M/L$.

\begin{figure}
\begin{center}
\includegraphics[bb= 50 90 560 740, angle=270, clip, width=\hsize]{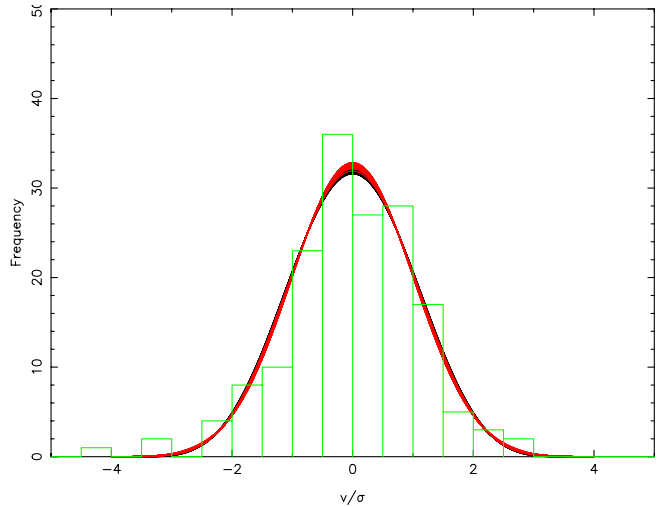}
\end{center}
\caption{The line of sight velocity distributions of the Michie model \#17 that fits the cluster are shown for each radial position at which we have a kinematic measurement. The inner half of the sample is shown in black, the outer half in red. These distributions are shown re-scaled so that the root mean square (r.m.s.) dispersion of the distributions is unity, which highlights the fact that they are very similar throughout the radial range of our data. The observed velocity distribution is also displayed (green histogram), with each point re-scaled by the r.m.s. of the corresponding model. Both the observed and best-fitting model distributions suggest that Gaussian line of sight velocity distributions are required.}
\label{fig:gaussiantest}
\end{figure}

Before attempting any further modelling it will be helpful to assess the shape of the velocity distribution that the data display. The difficulty with this is that the amount of kinematic data is rather limited, so it is not very informative to show directly the observed line of sight velocity distribution at various radii. So instead in Fig.~\ref{fig:gaussiantest}, we show the predicted line of sight velocity distributions of the Michie model \#17 at the radii of the kinematic observations. The distributions for the inner half of the kinematic data are shown in black, the outer half in red. These distributions have been scaled to have r.m.s. dispersion of unity. Note that despite the fact that the model has a strongly varying anisotropy (see Fig.~\ref{fig:betamodel17}), the integration along the line of sight yields distributions of very similar Gaussian shape at each radius. The green histogram shows the distribution of observed radial velocities, where each datum is divided by the r.m.s. dispersion of the line of sight velocity distribution of the Michie model at the corresponding radius. The data clearly follow closely a Gaussian distribution (obviously the precise statistic depends on how one deals with outliers, but as we have shown above, the Michie model \#17 is an excellent fit to the data). Any model that we consider has to conform to this fact.

\subsection{MCMC Jeans equation analysis}

The highly radially-anisotropic toy model shown in Fig.~\ref{fig:toymodels} (turquoise line), suggests that a more general model may help to relieve the pressure on MOND set by our data. The option we choose to explore next is to use the Jeans equation to determine profiles of the radial and tangential velocity dispersion that are consistent with the density profile and potential. The Jeans equation for a spherically-symmetric system is (\citealt{Binney:1987p11821}, Eqn 4-55):
\begin{equation}
{{G M(r)}\over{r }} = - \overline{v_r^2} \Bigg[ {{\mathrm{d} \ln \rho}\over{\mathrm{d} \ln
        r}} + {{\mathrm{d} \ln \overline{v_r^2}}\over{\mathrm{d} \ln r}} + 2\Big(1 -
    {{\overline{v_\theta^2}}\over{\overline{v_r^2}}} \Big) \Bigg] \, ,
\label{eqn:Jeans}
\end{equation}
where $M(r)$ is the cumulative mass inside a radius $r$.

We assume that the density profile conforms accurately to that of the Michie model \#17, and fix the potential by selecting a value for the mass to light ratio. We choose to simulate the set $(M/L)_V=\{1.6, 1.8, \dots, 3\}$, which span the astrophysically-plausible range according to the evolutionary synthesis models cited above. The challenge is to find now functions $\overline{v_r^2}(r)$ and $\overline{v_\theta^2}(r)$ that are allowed by Eqn.~\ref{eqn:Jeans}, and are also consistent with the observed data. The approach we take is to employ a Markov-Chain Monte Carlo (MCMC) scheme to fit trial distributions of $\overline{v_\theta^2}(r)$, in a similar (but not identical) fashion to \citet{Ibata:2009p13142}. We parametrise $\overline{v_r^2}(r)$ at 7 radial locations, and fit a cubic spline to these points, with the constraints that $\overline{v_r^2}$ is positive and falls to zero at the tidal radius. Starting from an initial guess for the 6 free parameters, the algorithm solves for $\overline{v_\theta^2}(r)$, and uses the density profile to project these quantities onto the line of sight at each radial point for which we have kinematic data. These line of sight velocity distributions are then convolved with the uncertainty distribution appropriate for each data point. A binary fraction of zero is assumed. The likelihood of the trial model is computed from Eqn.~\ref{eqn:likelihood}. 

\begin{figure}
\begin{center}
\includegraphics[width=\hsize]{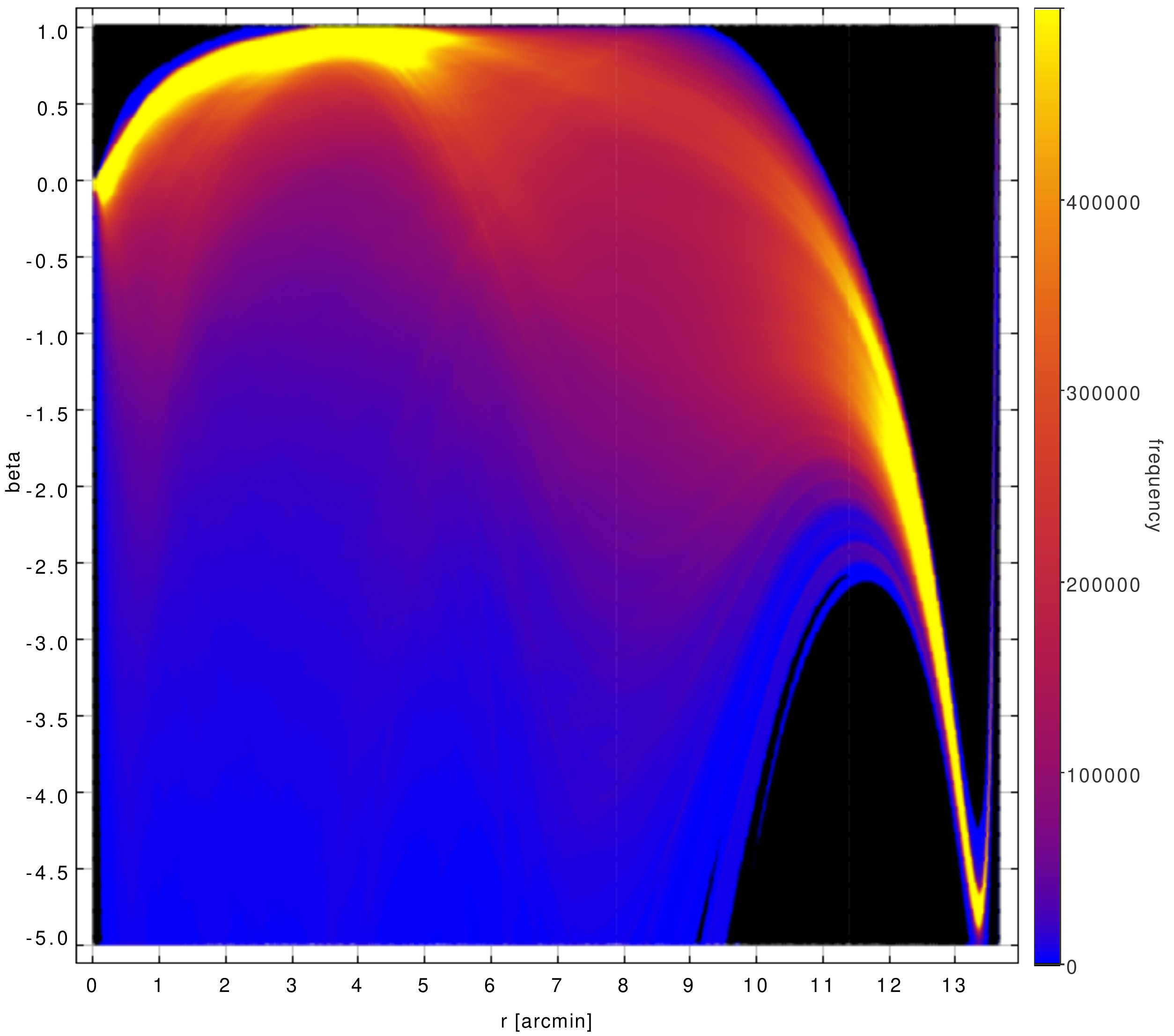}
\end{center}
\caption{The density of solutions of $\beta(r)$ derived using the Jeans equation in a Markov-Chain Monte Carlo simulation with $10^6$ steps that was implemented to fit the observed velocity data. The underlying density model is that of the Michie model \#17, with a MOND potential calculated taking $(M/L)_V=1.6$. Note that most of these solutions likely do not correspond to dynamically-possible models.}
\label{fig:MCMC}
\end{figure}

The MCMC algorithm we developed to sample the posterior probability distribution function is a new hybrid scheme which we intend to detail in a forthcoming contribution. To sample the full parameter space the algorithm employs Parallel Tempering \citep{Gregory:2005p11863}, a technique similar to simulated tempering, which uses parallel chains to probe the likelihood surface at different ``temperatures'' (a higher temperature produces  a shallower likelihood surface, allowing the algorithm to reach distant regions beyond a local maximum). However, instead of stepping though parameter space with the traditional Metropolis-Hastings criterion, we use the Affine-Invariant Ensemble Sampler of \citet{Goodman:2010p15978}. The \citet{Goodman:2010p15978} algorithm with ``stretch'' moves is very convenient as it naturally adapts the proposal step size (and, although not of paramount importance to the present problem, the affine-invariance property ensures that one does not need to know the relative scales of the parameters in advance). Tests show that our new hybrid scheme is very powerful, outperforming Parallel Tempering or the \citet{Goodman:2010p15978} algorithm on a variety of computationally-challenging problems.

\begin{figure}
\begin{center}
\includegraphics[bb= 50 90 560 740, angle=270, clip, width=\hsize]{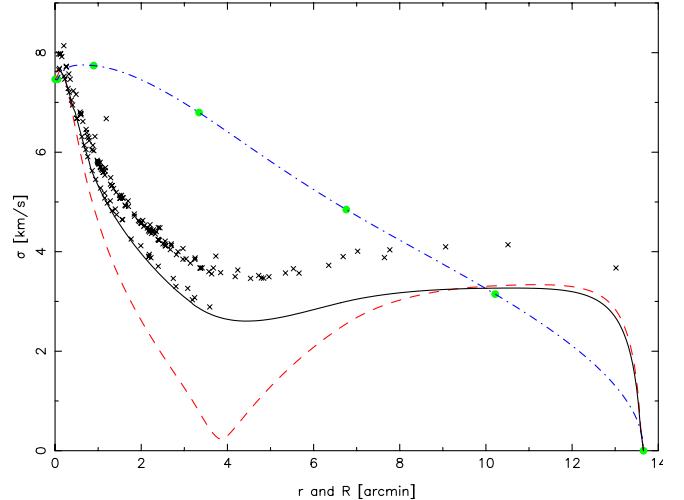}
\end{center}
\caption{The velocity dispersion profiles of the most likely model in MOND resulting from the Jeans equation solutions shown in Fig.~\ref{fig:MCMC}. The diagram demonstrates the working of the fitting procedure: the large dots mark the 7 points (the two inner points are close and appear merged) that were used to anchor the spline fit (dot-dashed line) to $\sigma_r(r)$. Assuming the fit to the luminosity profile made for model \#17, and with $(M/L)_V=1.6$, the Jeans equation is used to recover the $\sigma_\theta(r)$ profile (dashed line). Projecting these along the line of sight results in $\sigma_{v}(R)$ (continuous line). The crosses mark $\sigma_{v}$ at the positions of the kinematic measurements after convolving with the corresponding observational uncertainties. A visual comparison to Fig.~\ref{fig:vel_radius} shows clearly that even this model, which is the most likely of the MOND solutions, over-predicts the velocity dispersion at large radius.}
\label{fig:best_fit_MCMC}
\end{figure}

The particular implementation we constructed has 6 parallel MCMC chains, each sampled with an ensemble of 100 points per iteration. The chains are swapped (i.e. ``temperered'') once every 100 iterations on average. We add two astrophysical priors to the solution: first, the solution must conform to the GDSAI (Eqn.~\ref{eqn:GDSAI}), and second, we require that $\xi_{half,s} < 1.5$, which as discussed in \S~\ref{sec:stability}, avoids the most extremely radially unstable configurations. We further require that $\overline{v_\theta^2}(r)> 0$, which for practical (technical) reasons we implement as a very strong prior. Since we have found that the observed velocity distribution is closely Gaussian (Fig.~\ref{fig:gaussiantest}), we take $\overline{v_r ^2}(r)$ and $\overline{v_\theta^2}(r)$ to be truncated Gaussians. The truncation velocity at $r$ is set to $\sqrt{2\Psi(r)}$, where $\Psi(r)$ is the local relative potential (see \citealt{Binney:1987p11821}).

Figure~\ref{fig:MCMC} shows $10^6$ steps of the Markov Chain running in a MOND potential with $(M/L)_V=1.6$, the mass to light value that gives the highest likelihood of all the simulations. The $\beta$ parameter is calculated as a function of (non-projected) radius $r$ from spline fits to the radial and tangential velocity dispersion profiles. The density of solutions in the Markov Chain is proportional to their likelihood, so it is clear from this figure that the data --- even in MOND --- favor an anisotropy profile that is similar to a Michie model in the central regions, being isotropic in the very center and mostly radially anisotropic out to $\approx 5\arcmin$. Between $\approx 5\arcmin$ and $\approx 11\arcmin$ the solutions are highly degenerate and the data do not appear to favor any particular model. We have checked that $10^6$ iterations are sufficient for convergence: repeated MCMC simulations with very different initial parameters recover identical best-fit parameters to better than $0.3\%$. We are aware that our choice of anchoring $\overline{v_\theta^2}(r)$ at 7 radial locations (rather than some other number, and at some other locations), as well as the choice of employing cubic spline interpolation, and the implementation of the priors, likely have a significant impact on the probability density function of solutions found in Fig.~\ref{eqn:Jeans}. However, we will not explore this complex issue further, as our aim here is only to show that possible alternative solutions may exist.

We stress that the Jeans equation analysis does not guarantee that the solutions are physically plausible, indeed most of the anisotropy profiles in Fig.~\ref{fig:MCMC} probably do not correspond to self-consistent models (i.e., models with positive distribution function, which are a steady-state solution of the collisionless Bolzmann equation). Further work is required to check the solutions; and clearly given the number of possibilities this is a monumental task!  With this caveat in mind, we note that the {\it kinematics} of the most likely MOND solution from the MCMC analysis, whose velocity dispersion profiles are displayed in Fig.~\ref{fig:best_fit_MCMC}, is a factor of 350 less likely than the most likely solution in an identical MCMC simulation undertaken in Newtonian gravity. Although further work is required to rule out MOND categorically with the kinematics of this cluster, the present data already challenge the theory severely.

\subsection{External field effect}
\label{sec:efe}

\begin{figure}
\begin{center}
\includegraphics[width=\hsize]{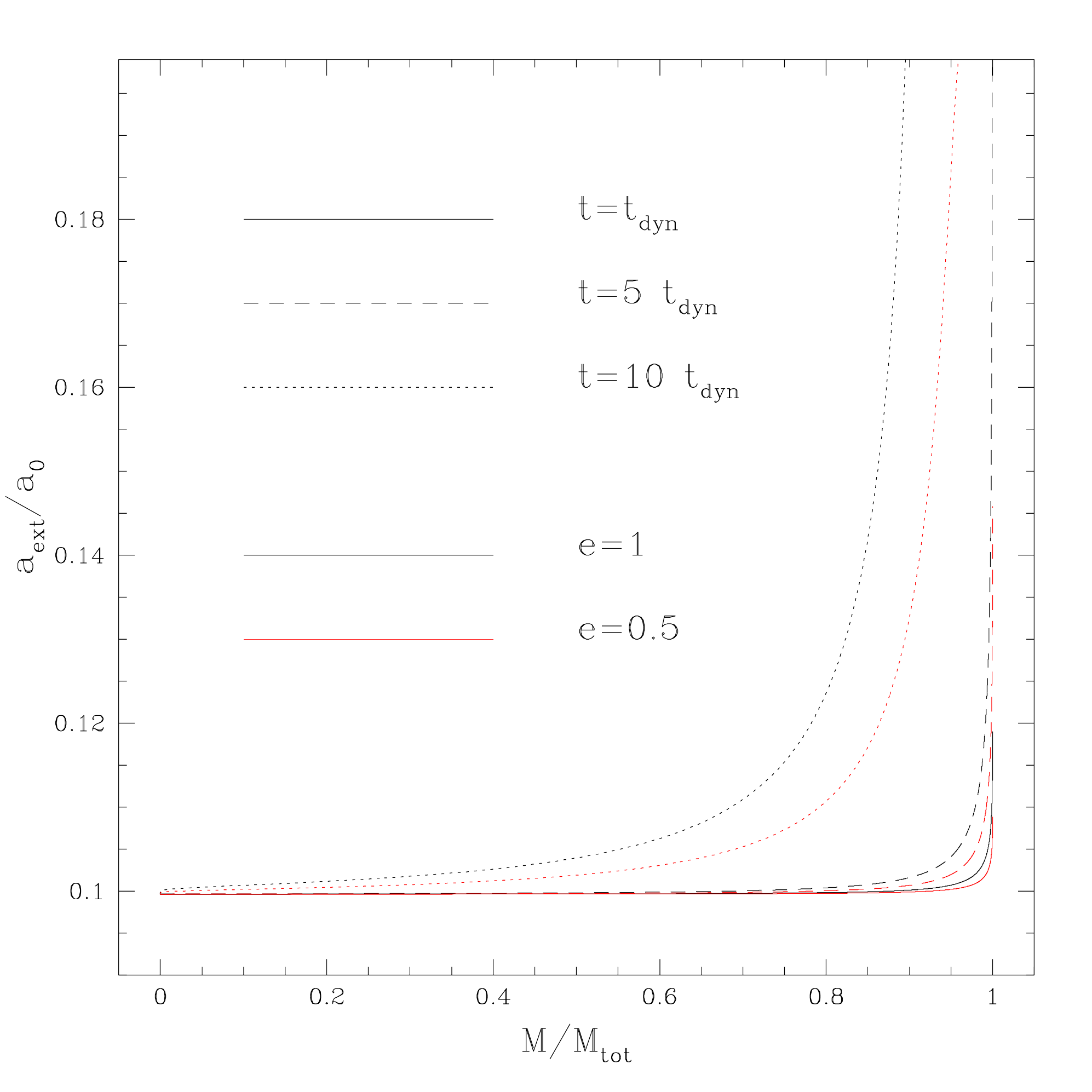}
\end{center}
\caption{External acceleration exerted by the Milky Way on NGC~2419 after a time interval $t=1$, 5 and 10 $t_{dyn}(r)$ as a function of the mass fraction comprised within $r$. The case of the two different eccentric orbits $e=0.5$ and $e=1$ are shown with red and black lines, respectively.}
\label{fig:orb}
\end{figure}

Throughout the paper we have treated the globular cluster NGC~2419 as isolated. This is justified by the large distance of the cluster from the Galactic center, suggesting that tidal effects are small and that, in MOND, the EFE is not important. However, given the poor performance of MOND at reproducing the kinematics of this object, it is worth trying to estimate quantitatively the EFE and to verify whether the discrepancy between data and models can be accounted for by the EFE. As pointed out in the Introduction, the gravitational field of the Galaxy at the location of NGC~2419 is estimated to be, in modulus, $g_{ext}\sim0.1 a_0$ and directed approximately towards the Galactic center\footnote{Independent of the gravity law, a cluster located in the plane of the Galaxy at a distance $R$ from the Galactic center experiences a gravitational field $\propto v^2_{rot}(R)/R$ directed towards the Galactic centre (where $v_{rot}$ is the rotational velocity of the Galaxy).  If the cluster is out of the plane, the field it experiences could be different in MOND and in Newtonian gravity with dark matter, due to the component of the field orthogonal to the plane \citep[][]{Nipoti:2007p15573}. However, this effect should not be strong for NGC~2419, given its large Galactocentric distance and relatively small distance from the Galactic plane.}. In the frame of reference of the cluster, this external field will change in time, both in direction and in magnitude, as a consequence of the orbital motion of the cluster.  The effect of such a variation on the internal kinematics of the cluster is linked to the ratio between the internal dynamical time ($t_{dyn}$) and the orbital period ($P_{orb}$). Although the orbit of NGC~2419 is unknown, we can set an upper limit to this ratio by assuming the cluster is presently at its apocenter.  As a first order approximation, we calculated the orbital period assuming planar orbits within a spherical isothermal Galactic potential $\phi_{MW}=v_{circ}^{2} \ln R_{GC}$ with $v_{circ}=187 \kms$ (to have $g_{ext}=0.1~a_{0}$ at $R_{GC}=94.8 \kpc$).  The dynamical time is calculated as
$$t_{dyn}(r)=2 \pi\sqrt{\frac{r}{g_{int}(r)}} \, ,$$ 
and we adopt the values of $g_{int}(r)$ predicted by the best MOND model \#24.  Following the above considerations, even assuming a free falling orbit (e=1) we obtain $t_{dyn}/P_{orb}=0.006$ at the half-mass radius and $t_{dyn}/P_{orb}=0.123$ at the tidal radius.  To understand the effect of such a variation on the internal cluster kinematics, we plot in Fig. \ref{fig:orb} the external acceleration exerted on the cluster after a time $t$ as a function of the fraction of mass comprised within the radius where $t/t_{dyn}(r)=$1, 5 and 10, for two different eccentric orbits ($e=0.5$ and 1).  It is evident that in all cases more than 90\% of the cluster feels an almost constant external acceleration (always smaller than $0.2~a_{0}$) after several dynamical times.  So it is reasonable to assume that the cluster is in equilibrium in the MOND potential calculated with the present-day external field $-\vec \del \varphi_{ext}$ (assumed uniform).  Such an equilibrium configuration will not be spherically symmetric and must be a solution of Equation~(\ref{eq_mond}) with boundary conditions $\vec\del \varphi \to \vec\del \varphi_{ext}$ for $|{\vec r}|\to \infty$ and density distribution $\rho$ such that, in projection, it reproduces the observed surface-brightness profile of NGC~2419.

\begin{figure}
\begin{center}
\includegraphics[width=\hsize]{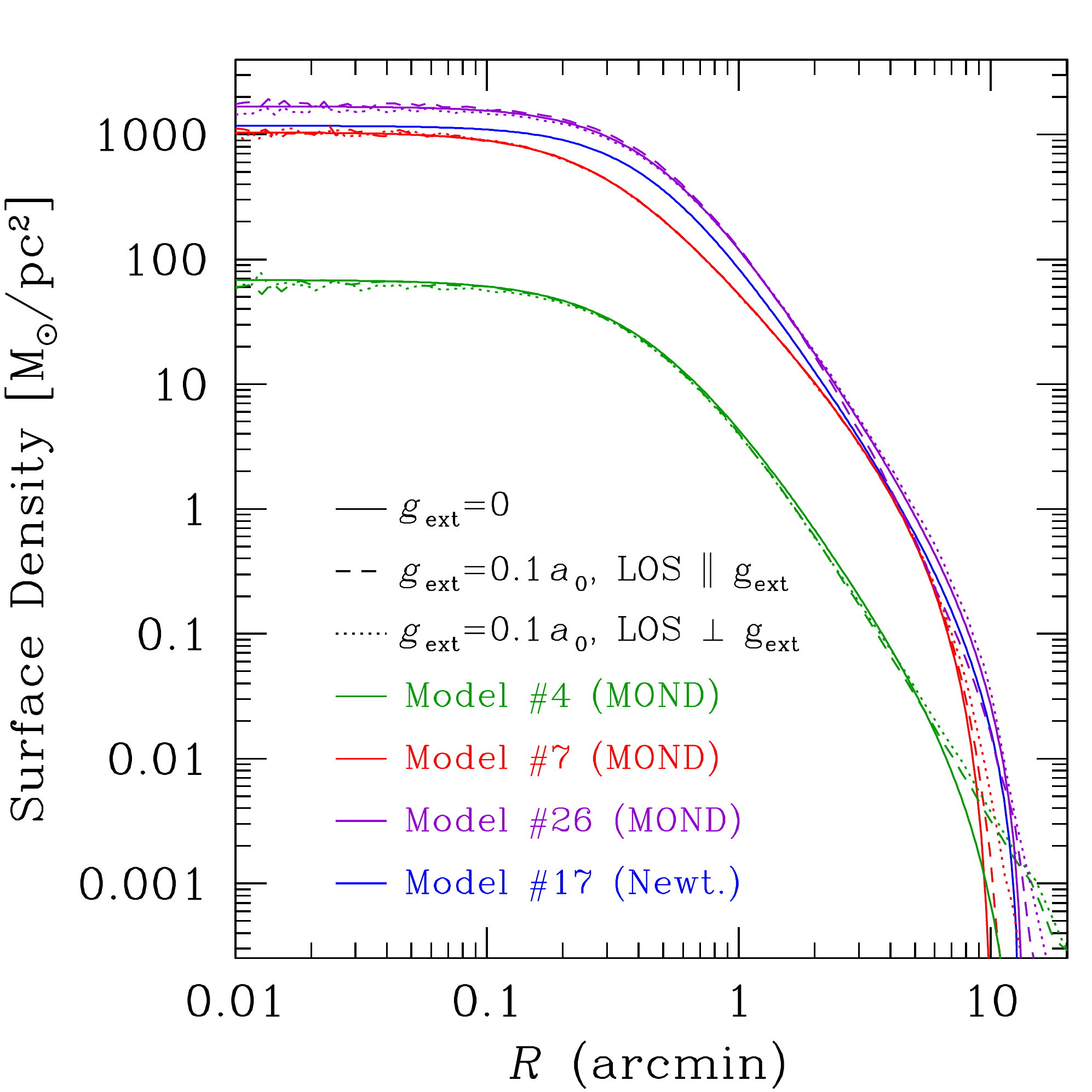}
\end{center}
\caption{Surface-density profiles of MOND models of NGC~2419 in the presence of the external field of the Galaxy and, for comparison, of the best-fitting Newtonian model \#17.}
\label{fig:sbefe}
\end{figure}

\begin{figure}
\begin{center}
\includegraphics[width=\hsize]{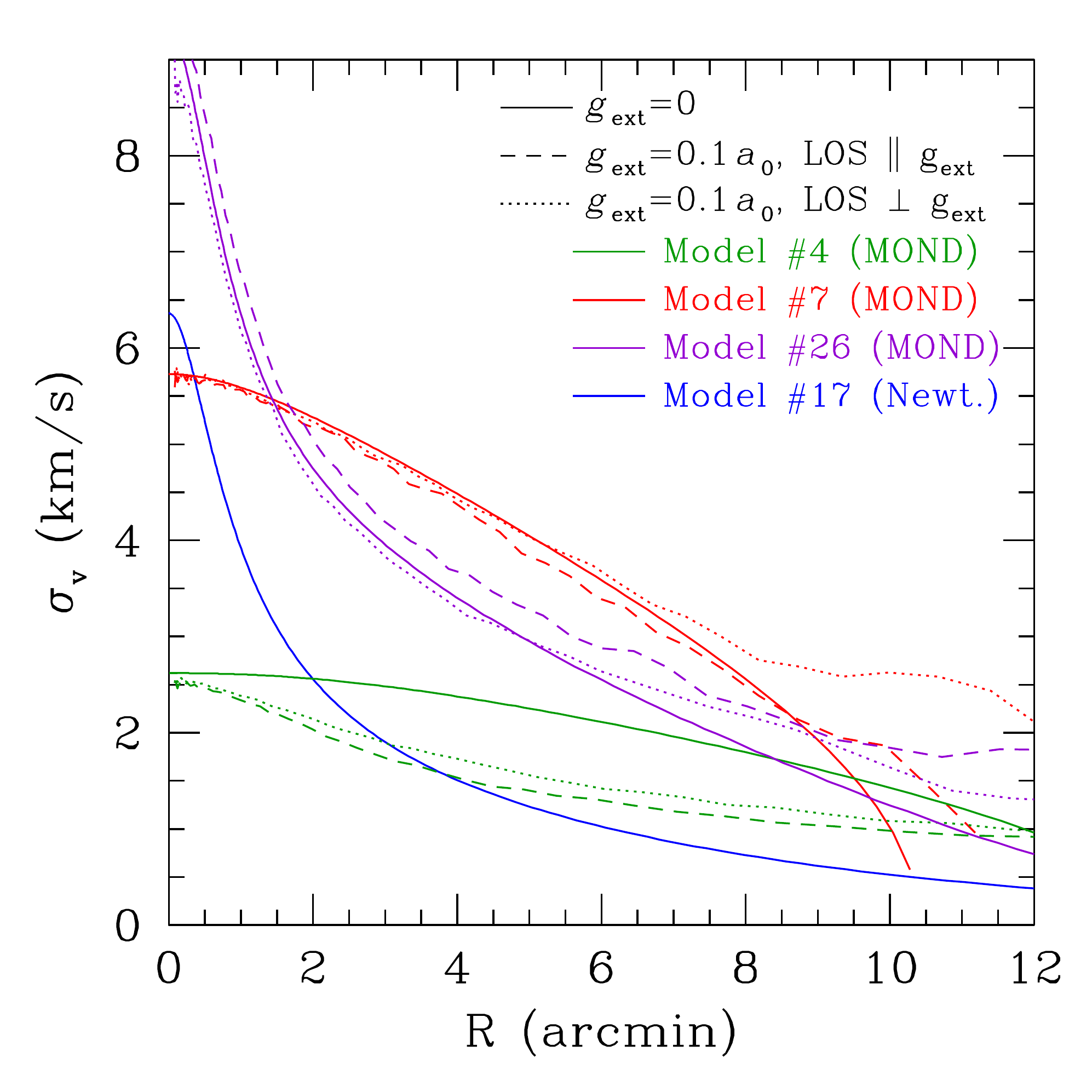}
\end{center}
\caption{Velocity-dispersion profiles of MOND models of NGC~2419 in  the presence of the external field of the Galaxy and, for comparison, of the best-fitting Newtonian model \#17.}
\label{fig:efe}
\end{figure}

We try to build such equilibrium models using the MOND N-body code \NMODY\ (see Section~\ref{sec:stability}), in which we implemented the EFE by imposing the above boundary conditions. In practice, given a density distribution $\rho$, the MOND field is calculated by starting from a guessed potential with the correct asymptotic behavior ($\vec\del \varphi \to \vec\del \varphi_{ext}$) and then the solution $\vec\del \varphi$ is found with the same relaxation procedure used in the absence of the external field \citep[see][]{Londrillo:2009p15575}. Of course, the particles are then evolved in phase-space using only the ``internal part'' of the acceleration $-(\vec\del \varphi-\vec\del \varphi_{ext})$.  While in the absence of the external field we were able to set up equilibrium initial N-body models using the distribution function~(\ref{eq_df}), in the presence of the external field we cannot do better than initializing our N-body simulations with quasi-equilibrium distributions. In particular, exploiting the fact that $g_{ext}\ll a_0$, we simply use N-body realizations (with $8\times10^5$ particles) of the spherically symmetric models of NGC~2419 reported in Table~\ref{tab:models} and let them evolve in the presence of the external field. We find that the models rapidly reach a new equilibrium configuration, which is axisymmetric with symmetry axis along the direction of the external field. In Figs.~\ref{fig:sbefe} and \ref{fig:efe} we plot, respectively, the circularized surface-density profiles and circularized line-of-sight velocity-dispersion profiles of the end-products of simulations with $g_{ext}=0.1 a_0$, in which the initial conditions correspond to MOND models \#4 (the best-fitting among the isotropic models, with $(M/L)_V=0.1$), \#7 (an isotropic model with more realistic $(M/L)_V=1.346$) and \#26 (the radially anisotropic model which has the highest likelihood among all the considered MOND models fit to sample A). For each model we plot the profiles with line-of-sight both parallel and orthogonal to the external field; for comparison we also plot the profiles of the same MOND models in the absence of the external field as well as that of the best-fitting Newtonian model \#17.  We note that the end-products are not highly flattened: in particular, the ellipticity is in all cases $\epsilon\lsim 0.01$ when the cluster is viewed along the direction of the external field (which should be approximately the case for our observations of NGC~2419) and $\epsilon\lsim 0.15$ when the line-of-sight is orthogonal to $\gvext$.  From Fig.~\ref{fig:sbefe} it is clear that the EFE has little influence on the surface brightness profiles: only the very low-mass model \#4 presents a significant deviation from the reference model \#17, but only at $R\gsim 7$~arcmin.  Similarly, the velocity dispersion profiles of the end-products of the EFE simulations do not deviate substantially from the profiles of the same models with no external field (see Fig.~\ref{fig:efe}); again, the model most affected by the EFE is the $(M/L)_V$=0.1 model \#4, with a significantly lower velocity dispersion even at $R\lsim 5$ arcmin, where we have most spectroscopic data.  Overall, a comparison of the MOND models with EFE with the best-fitting Newtonian model \#17 (blue curve in Fig.~\ref{fig:efe}) clearly shows that an external field of magnitude $0.1 a_0$ does not help reconcile MOND models with the data.  We ran similar simulations with an external field $g_{ext}=0.2 a_0$, finding slightly stronger effects, but the same qualitative behavior.  We conclude that for NGC~2419 the EFE is unlikely to represent the way out for MOND.

\section{Conclusions}
\label{sec:conclusions}

We present new spectroscopic data of the distant globular cluster NGC~2419, the most distant of the large Galactic globular clusters. Taken with the DEIMOS instrument on the Keck II telescope, these  have been reduced with a powerful new spectral extraction package that was designed partly to take advantage of these state-of-the-art data of NGC~2419. By avoiding pixel resampling we are able to maintain optimal accuracy, and we avoid having correlated noise. The algorithm has yielded an accurate sample of cluster kinematics with reliable velocity uncertainties (derived by an MCMC fit to the spectra), which have been tested against the external dataset presented by B09. The spectroscopic sample contains 166 likely cluster member stars. We have also reanalyzed deep and wide-field photometric star-counts surveys of the cluster, taken with HST/ACS and Subaru/SuprimeCam.

A set of 26 Newtonian and MOND Michie models of the cluster were constructed to compare against the observations, and N-body realizations of these models were evolved to ensure dynamical stability. The Michie models were projected into velocity distributions along the line of sight, convolved with the velocity uncertainties and a plausible model for binaries, allowing us to calculate the likelihood of each kinematic data point for each model. We also fold in the likelihood of the star-counts profile, carefully taking into account the footprint of the imaging surveys. Due to the easy scalability in the Newtonian case, it was readily feasible to scan a large range of $(M/L)_V$ values. For MOND, we started from plausible initial values and homed-in manually on the best-fitting $(M/L)_V$. 

We find that isotropic Michie models in both theories of gravity are very much at odds with the observations, and can be ruled out. However, our anisotropic Newtonian model with $r_a/r_c=1.5$ and $(M/L)_V=1.903$, and with total mass $M=9.12\times10^5\msun$ is in good statistical agreement with the present observations, and is therefore an excellent representation of the system. We note, parenthetically, that this implies that the structure and dynamics of the cluster can be reproduced satisfactorily without invoking dark matter, in agreement with B09. In contrast, the best anisotropic MOND model is a factor of $\sim 40000$ less likely.

To extend these conclusions to more general models, we implemented a Markov-Chain Monte-Carlo sampling routine to probe solutions to the Jeans Equation that are consistent with the luminosity profile and kinematics. While such a procedure does not ensure physically-consistent solutions (with a non-negative distribution function), it nevertheless suggests that any plausible MOND model will have severe difficulties reproducing the data, as the best of the MOND solutions is still a factor of $350$ less likely than the best Newtonian solution.

Finally, we show with N-body simulations that the external acceleration field is unlikely to affect significantly the above conclusions.

Until now, most of the existing analyses in the low external field regime focussed on the global velocity dispersion in sparse clusters \citep[e.g.][]{Haghi:2009p16438, Haghi:2011p15956}. In contrast,  here we make use of the full velocity distribution and its radial dependence, thus making our test much less dependent on the adopted $(M/L)_V$. Furthermore, those authors who have studied cluster radial velocities, generally bin their kinematic data to measure first and second Gaussian moments. The likelihood analysis presented here works directly on the individual velocity measurements for each star, thus conserving the full spatial resolution, and the full discriminative power of the kinematic data.

Our conclusions rest on a number of assumptions:
\begin{itemize}
\item The cluster is taken to be a spherical body; while we have shown that the kinematic sample does not display significant evidence for a preferred axis of rotation, there is marginal evidence for oblateness in the integrated surface brightness distribution at about the $2\sigma$ confidence level in the inner $2\arcmin$ (B07). At larger radii the cluster appears rounder. If the cluster in reality departs significantly from spherical symmetry, the present analysis will not be valid. Orbits within a triaxial, or indeed a more complex structure, will in general be very different to the spherical case that was modelled. Since we observe the cluster in projection, and have only limited constraints on the line of sight depth, it is possible that the cluster is elongated or compressed along the line of sight. If the cluster is undergoing tidal disruption, the natural expectation is that the structure should be elongated in this direction (given that satellite orbits are predominantly radial). The precise kinematic behavior is hard to predict. If a substantial number of unbound extra-tidal stars are present in the sample, the configuration would be similar to the suggestion by \citet{Kroupa:1997p16018} regarding dwarf satellite galaxies, and this would have the tendency to aument the apparent velocity dispersion, making the situation worse for MOND. In contrast, if the bound cluster stars have been stretched along the line of sight, conservation of phase space density suggests that the line of sight velocity dispersion should drop with respect to the spherical case. This might provide a get-out clause for MOND. While it is currently impossible to determine the three-dimensional shape of NGC~2419, globular clusters in nearby galaxies such as M31 (e.g. in the outer halo sample of \citealt{Mackey:2010p16135}) do not, to our knowledge, display significant morphological distortions in the direction pointing towards their host galaxy\footnote{As we have shown in \S\ref{sec:Introduction}, in the Milky Way the only remote cluster that is also massive (and has many stars to trace the velocity profile) is NGC2419. This is not the case in M31 where there are several clusters of this class, and in the ELT era it will be feasible to repeat the present analysis for many of them. If similar results are found, it will be difficult to invoke special alignment conditions.}.
\item We assume the cluster to be an isolated system, with a single self-consistent dynamical tracer population. At first sight this appears a poor assumption, since NGC~2419 has been proposed at various times to be the remnant of a once nucleated galaxy \citep{Mackey:2005p16264}, or a cluster ejected from the Sagittarius stream \citep{Newberg:2003p16424}, or even the Virgo stream \citep{CasettiDinescu:2009p16267}. While these claims remain somewhat conjectural at present, what concerns us most for the present discussion is whether the previous evolutionary history of this structure has possibly implanted dynamical contaminants that may invalidate our analysis. Clearly NGC~2419 does not possess the complex dynamical mixture observed in M54 at the heart of the Sagittarius dwarf galaxy \citep{Bellazzini:2008p8100,Ibata:2009p13142}, so the effect is not overwhelming, but is likely to be subtle. Furthermore, from the analysis by B09 and from the discussion above, it appears that there is probably no significant amount of dark matter bound to the cluster (although presumably a cored mini-halo with a large core radius could be accommodated). Thus although it is difficult to rule out the possibility that our analysis is affected by the complexities of the real system with its (possible) stream and its (possible) dark matter halo, we estimate that these issues are not significant. On the other hand, there is ample evidence \citep[][see also Fig.~\ref{fig:data} above]{Cohen:2010p16255} for a small ($\sim 0.2$~dex) spread in metallicity, indicating that we are not dealing with a single stellar population. Unfortunately, the kinematic data are not sufficiently numerous at present to attempt a more refined analysis, for instance, repeating the calculations presented above for several sub-samples split as a function of metallicity. However, inspection of Fig.~\ref{fig:data}d suggests that there is little radial variation of the metallicity distribution function with radius, so it would be very surprising if such an analysis changed significantly the present conclusions.
\item We assume that NGC~2419 is a static structure. This need not be the case if the cluster moves on a highly radial orbit. Indeed if it was once associated to the Sagittarius dwarf or the progenitor of the Virgo Stellar Stream, it is plausible that it has a similar pericenter to those structures ($\sim 10\kpc$). In that case, the cluster will have experienced a periodic, highly varying, external acceleration field. Although the consequences of this possibility cannot be fully fathomed without undertaking detailed N-body simulations in a realistic Galactic potential with a realistic orbit, we note that the dynamical time estimated above suggests that the cluster settles down into equilibrium on a short timescale compared to its orbital period. Thus for the present purposes of testing gravity, NGC~2419 should be effectively in dynamical equilibrium.
\item Implicit in our work is also the assumption of a constant $M/L$ throughout the cluster. Dark remnants such as black holes or neutron stars are likely to be unimportant in this context, since they contribute only 3\% and 0.8\% to the total mass budget, respectively, according to the BASTI models \citep{Percival:2009p17086}. However, faint main-sequence stars may be more problematic. If NGC~2419 experienced strong primordial mass-segregation, such as is inferred in nearby young starburst clusters \citep{Stolte:2005p17927, Stolte:2006p17879}, this would influence the line of sight velocity dispersion profile, lowering the central value and producing a shallower slope. We are currently attempting to measure the mass function in NGC2419 using HST star counts with the Wide Field and Planetary Camera 3 to address this problem directly (Dalessandro et al., in preparation); however, as discussed above, the similarity of the Blue Straggler Star profile to that of other populations already appears to exclude significant mass-segregation in this cluster (D08).
\end{itemize}

The present observations and analysis provide a strong test of the nature of gravity in the low acceleration regime. With the above assumptions and caveats, we conclude that the Newtonian approximation to General Relativity provides --- by far --- a better representation for the structure and dynamics of NGC~2419 than MOND.

\acknowledgments

R.I. gratefully acknowledges support from the Agence Nationale de la Recherche though the grant POMMME (ANR 09-BLAN-0228). We acknowledge the CINECA Award N. HP10C2TBYB, 2011 for the availability of high performance computing resources and support. C.N. is supported by the MIUR grant PRIN2008. M.B. acknowledges support by INAF through the PRIN-INAF 2009 grant CRA 1.06.12.10 (PI: R. Gratton) and by ASI through contracts COFIS ASI-INAF I/016/07/0 and ASI-INAF I/009/10/0.

\bibliography{N2419}
\bibliographystyle{apj}

\begin{table*}
\begin{center}
\caption{Spectroscopic measurements, with the corresponding photometry.}
\label{tab:data}
\begin{tabular}{ccccccccccc}
\tableline\tableline
star & mask/ID & RA      & Dec     & $ v$     & $\delta v$ & ${\rm I}$ & ${\rm V-I}$ & ${\rm [Fe/H]}$ & $R$      & Multiple \\
       &         & (J2000) & (J2000) & $(\kms)$ & $(\kms)$   & (mag)     & (mag)       &                & (arcmin) &  observations?        \\ 
\tableline
  1 & mask 1        &  7 38 08.90 & 38 52 53.1 & -21.506  &  2.277 &  17.404 &   1.176 &  -1.862 &   0.082 & n\\
  2 & B09  S31      &  7 38 08.64 & 38 53 00.0 & -23.690  &  0.430 &  16.315 &   1.339 &  -3.412 &   0.089 & n\\
  3 & B09  S17      &  7 38 08.01 & 38 52 53.1 & -10.120  &  0.810 &  16.104 &   1.403 &  -3.451 &   0.102 & y\\
  3 & mask 2        &  7 38 08.00 & 38 52 53.2 &  -9.294  &  2.264 &  16.104 &   1.403 &  -1.848 &   0.103 & y\\
  4 & mask 1        &  7 38 08.29 & 38 53 01.5 & -12.013  &  2.270 &  16.941 &   1.239 &  -1.828 &   0.118 & n\\
  5 & mask 1        &  7 38 07.86 & 38 52 56.9 & -28.078  &  2.365 &  18.556 &   1.056 &  -1.772 &   0.131 & n\\
  6 & mask 1        &  7 38 08.60 & 38 53 04.3 & -19.134  &  2.269 &  16.837 &   1.269 &  -1.862 &   0.158 & n\\
  7 & mask 2        &  7 38 07.66 & 38 53 02.0 & -17.968  &  3.140 &  16.980 &   1.224 &  -1.455 &   0.203 & n\\
  8 & B09  S22      &  7 38 08.80 & 38 52 42.9 & -33.980  &  0.640 &  16.124 &   1.398 &  -3.447 &   0.208 & n\\
  9 & B09  S38      &  7 38 07.67 & 38 52 45.5 & -10.220  &  0.430 &  16.487 &   1.327 &  -3.369 &   0.226 & n\\
 10 & mask 1        &  7 38 07.19 & 38 52 57.5 & -22.569  &  2.282 &  17.362 &   1.163 &  -1.360 &   0.261 & n\\
 11 & mask 2        &  7 38 09.44 & 38 53 06.3 & -11.755  &  2.267 &  17.244 &   1.203 &  -1.550 &   0.262 & n\\
 12 & B09  S5       &  7 38 10.07 & 38 52 54.3 & -16.410  &  0.670 &  15.807 &   1.514 &  -3.501 &   0.304 & n\\
 13 & mask 1        &  7 38 09.95 & 38 53 04.3 & -14.820  &  2.319 &  18.308 &   0.960 &  -2.088 &   0.321 & n\\
 14 & B09  S15      &  7 38 09.69 & 38 52 40.9 & -16.790  &  0.580 &  15.924 &   1.461 &  -3.484 &   0.327 & n\\
 15 & mask 1        &  7 38 09.27 & 38 52 35.7 & -27.394  &  2.311 &  18.237 &   1.020 &  -1.890 &   0.353 & y\\
 15 & mask 2        &  7 38 09.27 & 38 52 35.7 & -29.969  &  2.289 &  18.237 &   1.020 &  -1.916 &   0.353 & y\\
 16 & B09  S4       &  7 38 07.71 & 38 53 15.7 & -24.270  &  0.660 &  15.778 &   1.484 &  -3.517 &   0.380 & n\\
 17 & B09  S9       &  7 38 10.18 & 38 53 09.6 & -15.950  &  0.560 &  15.923 &   1.466 &  -3.483 &   0.407 & n\\
 18 & mask 2        &  7 38 09.47 & 38 52 32.0 & -14.065  &  2.277 &  17.685 &   1.139 &  -1.473 &   0.425 & n\\
 19 & mask 2        &  7 38 10.42 & 38 52 35.9 & -22.231  &  2.482 &  19.701 &   0.853 &  -1.436 &   0.488 & n\\
 20 & B09  S41      &  7 38 06.32 & 38 52 38.7 & -29.270  &  0.440 &  16.413 &   1.324 &  -3.389 &   0.505 & n\\
 21 & B09  S26      &  7 38 05.90 & 38 52 52.0 & -27.170  &  0.530 &  16.230 &   1.339 &  -3.434 &   0.510 & y\\
 21 & mask 1        &  7 38 05.90 & 38 52 52.1 & -24.912  &  2.278 &  16.230 &   1.339 &  -1.660 &   0.510 & y\\
 22 & mask 1        &  7 38 11.30 & 38 53 07.1 & -20.373  &  2.285 &  17.798 &   1.105 &  -1.907 &   0.580 & n\\
 23 & mask 2        &  7 38 11.48 & 38 53 02.6 &  -5.148  &  2.272 &  17.370 &   1.146 &  -1.962 &   0.592 & n\\
 24 & mask 1        &  7 38 10.30 & 38 52 25.6 & -30.009  &  2.358 &  18.918 &   0.968 &  -1.848 &   0.600 & n\\
 25 & mask 2        &  7 38 06.42 & 38 53 21.5 & -19.325  &  2.278 &  17.542 &   1.152 &  -1.546 &   0.602 & n\\
 26 & B09  S14      &  7 38 11.58 & 38 53 05.8 & -12.220  &  0.610 &  15.940 &   1.440 &  -3.485 &   0.624 & n\\
 27 & mask 2        &  7 38 05.80 & 38 53 17.3 & -24.109  &  2.282 &  17.758 &   1.123 &  -1.860 &   0.646 & n\\
 28 & B09  S20      &  7 38 07.60 & 38 53 34.1 & -23.660  &  0.550 &  16.082 &   1.419 &  -3.453 &   0.677 & y\\
 28 & mask 2        &  7 38 07.60 & 38 53 34.2 & -19.461  &  2.264 &  16.082 &   1.419 &  -1.407 &   0.679 & y\\
 29 & B09  S6       &  7 38 06.89 & 38 53 33.5 & -16.240  &  0.710 &  15.939 &   1.455 &  -3.481 &   0.716 & y\\
 29 & mask 1        &  7 38 06.89 & 38 53 33.6 & -18.856  &  2.274 &  15.939 &   1.455 &  -1.952 &   0.718 & y\\
 30 & mask 2        &  7 38 12.20 & 38 52 51.3 & -26.590  &  2.383 &  19.287 &   0.965 &  -1.533 &   0.721 & n\\
 31 & mask 1        &  7 38 04.83 & 38 53 04.1 & -31.310  &  2.309 &  17.711 &   1.172 &  -1.696 &   0.732 & n\\
 32 & mask 1        &  7 38 11.00 & 38 53 29.7 & -17.014  &  2.282 &  17.135 &   1.211 &  -1.607 &   0.756 & n\\
 33 & B09  S23      &  7 38 09.57 & 38 53 38.7 & -22.810  &  0.580 &  16.113 &   1.421 &  -3.444 &   0.759 & n\\
 34 & mask 1        &  7 38 05.55 & 38 52 23.1 & -20.172  &  2.366 &  18.746 &   0.988 &  -1.712 &   0.783 & n\\
 35 & mask 2        &  7 38 05.78 & 38 52 19.7 & -24.320  &  2.267 &  16.887 &   1.259 &  -1.806 &   0.792 & n\\
 36 & mask 2        &  7 38 05.16 & 38 53 28.3 & -29.221  &  2.270 &  16.588 &   1.301 &  -1.495 &   0.857 & n\\
 37 & B09  S12      &  7 38 06.00 & 38 53 37.5 & -22.160  &  0.630 &  16.037 &   1.432 &  -3.461 &   0.862 & n\\
 38 & mask 1        &  7 38 04.49 & 38 53 18.7 & -20.409  &  2.589 &  19.865 &   0.903 &  -1.527 &   0.877 & n\\
 39 & mask 2        &  7 38 08.75 & 38 53 48.8 & -15.343  &  2.430 &  16.715 &   1.348 &  -1.869 &   0.900 & n\\
 40 & mask 1        &  7 38 13.06 & 38 53 11.1 & -24.377  &  3.169 &  17.468 &   1.195 &  -1.605 &   0.926 & n\\
 41 & B09  S2       &  7 38 09.90 & 38 52 00.7 & -15.560  &  0.530 &  15.635 &   1.645 &  -3.512 &   0.943 & n\\
 42 & mask 1        &  7 38 05.13 & 38 52 11.6 & -20.403  &  2.296 &  17.992 &   1.066 &  -1.914 &   0.977 & n\\
 43 & mask 2        &  7 38 05.30 & 38 53 40.7 & -27.474  &  2.274 &  17.167 &   1.205 &  -1.388 &   0.986 & n\\
 44 & mask 1        &  7 38 11.98 & 38 52 11.0 & -25.504  &  2.278 &  16.291 &   1.370 &  -1.525 &   0.996 & n\\
 45 & mask 1        &  7 38 06.56 & 38 51 59.1 & -25.409  &  2.282 &  16.895 &   1.256 &  -1.914 &   1.004 & y\\
 45 & mask 2        &  7 38 06.56 & 38 51 59.1 & -24.024  &  2.263 &  16.895 &   1.256 &  -1.839 &   1.004 & y\\
 46 & mask 2        &  7 38 06.24 & 38 51 59.8 & -21.484  &  2.272 &  17.660 &   1.163 &  -1.913 &   1.019 & n\\
 47 & mask 2        &  7 38 12.77 & 38 52 16.6 & -23.786  &  2.480 &  19.785 &   0.966 &  -1.568 &   1.046 & n\\
 48 & mask 1        &  7 38 07.52 & 38 51 52.9 & -21.132  &  2.281 &  17.721 &   1.064 &  -2.055 &   1.051 & n\\
 49 & mask 1        &  7 38 12.31 & 38 52 09.7 & -26.248  &  2.282 &  17.506 &   1.164 &  -1.553 &   1.056 & n\\
 50 & B09  4        &  7 38 05.94 & 38 53 50.8 & -23.490  &  0.830 &  16.673 &   1.295 &  -3.327 &   1.057 & n\\
 51 & mask 1        &  7 38 03.72 & 38 53 24.9 & -19.591  &  2.301 &  16.892 &   1.239 &  -1.846 &   1.058 & y\\
 51 & mask 2        &  7 38 03.72 & 38 53 24.9 & -19.566  &  2.274 &  16.892 &   1.239 &  -1.840 &   1.058 & y\\
 52 & mask 1        &  7 38 12.76 & 38 53 36.2 & -14.816  &  2.265 &  16.121 &   1.409 &  -1.705 &   1.076 & n\\
 53 & mask 1        &  7 38 13.87 & 38 52 34.2 & -20.229  &  2.308 &  18.633 &   1.049 &  -1.847 &   1.099 & n\\
 54 & mask 1        &  7 38 11.63 & 38 51 58.8 & -17.571  &  2.263 &  15.673 &   1.584 &  -1.989 &   1.115 & n\\
 55 & mask 1        &  7 38 04.16 & 38 52 10.7 & -19.954  &  2.290 &  17.628 &   1.165 &  -1.797 &   1.122 & n\\
 56 & B09  31       &  7 38 11.15 & 38 51 53.4 & -18.020  &  1.080 &  16.774 &   1.283 &  -3.303 &   1.147 & n\\
 57 & B09  S16      &  7 38 14.21 & 38 52 36.0 & -19.990  &  0.460 &  16.054 &   1.449 &  -3.452 &   1.153 & y\\
 57 & mask 1        &  7 38 14.21 & 38 52 36.1 & -22.037  &  2.360 &  16.054 &   1.449 &  -1.798 &   1.153 & y\\
 58 & mask 2        &  7 38 12.41 & 38 52 02.6 & -18.242  &  2.273 &  17.702 &   1.133 &  -1.865 &   1.156 & n\\
 59 & mask 2        &  7 38 06.42 & 38 51 49.7 & -28.054  &  2.492 &  19.735 &   0.940 &  -1.616 &   1.160 & n\\
 60 & mask 1        &  7 38 03.05 & 38 53 24.4 & -19.784  &  2.661 &  19.872 &   0.940 &  -1.443 &   1.171 & n\\
 61 & mask 2        &  7 38 03.75 & 38 53 39.6 & -19.113  &  4.424 &  17.557 &   1.163 &  -2.202 &   1.189 & n\\
 62 & B09  38       &  7 38 02.87 & 38 52 24.1 & -23.630  &  0.400 &  16.215 &   1.375 &  -3.429 &   1.212 & n\\
 63 & B09  34       &  7 38 04.90 & 38 51 50.6 & -20.960  &  1.280 &  16.937 &   1.235 &  -3.272 &   1.281 & n\\
 64 & mask 1        &  7 38 14.92 & 38 52 36.5 & -37.390  &  2.279 &  17.607 &   1.122 &  -1.956 &   1.285 & n\\
 65 & mask 1        &  7 38 03.37 & 38 52 06.0 & -23.196  &  2.582 &  19.713 &   0.944 &  -1.354 &   1.290 & n\\
 66 & mask 1        &  7 38 02.74 & 38 52 14.4 & -23.532  &  2.285 &  17.085 &   1.234 &  -1.666 &   1.310 & n\\
 67 & mask 2        &  7 38 01.69 & 38 52 54.2 & -16.781  &  2.343 &  18.806 &   0.979 &  -1.915 &   1.327 & n\\
 68 & B09  S3       &  7 38 01.92 & 38 53 15.4 & -26.520  &  0.610 &  15.783 &   1.492 &  -3.513 &   1.327 & n\\
 69 & mask 1        &  7 38 01.84 & 38 52 31.1 & -14.943  &  2.301 &  17.318 &   1.160 &  -1.950 &   1.357 & n\\
 70 & B09  36       &  7 38 03.29 & 38 51 57.2 & -20.900  &  1.290 &  17.074 &   1.233 &  -3.236 &   1.399 & n\\
\tableline
\end{tabular}
\tablecomments{For those stars that have multiple measurements, we adopt the measurement with the lowest uncertainty.}
\end{center}
\end{table*}

\begin{table*}
\begin{center}
\setcounter{table}{0}
\caption{...continued...}
\begin{tabular}{ccccccccccc}
\tableline
 71 & mask 1        &  7 38 14.59 & 38 53 41.9 & -24.363  &  2.301 &  17.694 &   1.139 &  -1.649 &   1.419 & n\\
 72 & B09  35       &  7 38 03.55 & 38 51 51.9 & -18.450  &  0.980 &  17.090 &   1.218 &  -3.236 &   1.426 & y\\
 72 & mask 2        &  7 38 03.54 & 38 51 52.0 & -18.874  &  2.269 &  17.090 &   1.218 &  -1.386 &   1.426 & y\\
 73 & mask 1        &  7 38 01.35 & 38 53 27.7 & -22.030  &  2.369 &  18.710 &   1.048 &  -1.801 &   1.497 & n\\
 74 & mask 2        &  7 38 00.77 & 38 52 58.1 & -23.904  &  2.280 &  17.388 &   1.200 &  -1.851 &   1.507 & n\\
 75 & mask 2        &  7 38 11.61 & 38 51 31.6 & -21.251  &  2.291 &  18.378 &   1.078 &  -1.713 &   1.514 & n\\
 76 & mask 1        &  7 38 15.89 & 38 53 25.4 & -21.291  &  2.363 &  18.980 &   1.022 &  -1.781 &   1.523 & n\\
 77 & mask 1        &  7 38 16.42 & 38 53 01.0 & -12.572  &  2.487 &  19.408 &   0.974 &  -1.571 &   1.543 & n\\
 78 & B09  7        &  7 38 11.93 & 38 54 19.4 & -22.400  &  1.260 &  17.877 &   1.112 &  -3.053 &   1.558 & n\\
 79 & mask 1        &  7 38 09.61 & 38 51 21.8 & -23.079  &  2.430 &  19.256 &   1.011 &  -1.690 &   1.566 & n\\
 80 & B09  1        &  7 38 02.83 & 38 54 01.7 & -21.630  &  1.310 &  17.461 &   1.145 &  -3.156 &   1.569 & n\\
 81 & mask 1        &  7 38 10.67 & 38 51 23.5 & -22.992  &  2.463 &  19.330 &   1.001 &  -1.487 &   1.580 & n\\
 82 & mask 2        &  7 38 16.70 & 38 52 36.1 & -21.921  &  2.262 &  16.886 &   1.253 &  -1.406 &   1.624 & n\\
 83 & mask 2        &  7 38 14.96 & 38 51 51.6 & -18.102  &  2.259 &  15.951 &   1.455 &  -1.856 &   1.640 & n\\
 84 & mask 1        &  7 38 13.39 & 38 54 19.2 & -26.983  &  2.586 &  19.627 &   0.985 &  -1.439 &   1.696 & n\\
 85 & mask 2        &  7 37 59.90 & 38 53 13.2 & -18.270  &  2.394 &  19.105 &   0.993 &  -1.634 &   1.703 & n\\
 86 & B09  S10      &  7 38 16.92 & 38 53 35.0 & -20.750  &  0.580 &  16.227 &   1.384 &  -3.423 &   1.768 & n\\
 87 & mask 2        &  7 37 59.14 & 38 53 09.3 & -17.711  &  2.358 &  18.715 &   1.044 &  -1.483 &   1.839 & n\\
 88 & mask 1        &  7 38 02.33 & 38 51 31.0 & -24.023  &  2.275 &  17.292 &   1.201 &  -1.966 &   1.845 & n\\
 89 & mask 1        &  7 38 17.46 & 38 52 04.5 & -18.671  &  2.266 &  17.229 &   1.180 &  -2.035 &   1.934 & n\\
 90 & mask 2        &  7 38 15.02 & 38 51 25.4 & -24.403  &  2.273 &  17.649 &   1.160 &  -1.505 &   1.957 & n\\
 91 & B09  24       &  7 38 18.18 & 38 52 15.9 & -21.680  &  1.110 &  17.544 &   1.158 &  -3.130 &   1.991 & n\\
 92 & mask 2        &  7 38 15.58 & 38 51 28.4 & -15.183  &  2.305 &  18.725 &   0.977 &  -1.917 &   1.993 & n\\
 93 & mask 1        &  7 38 15.59 & 38 54 22.0 & -18.895  &  2.293 &  18.275 &   1.083 &  -1.837 &   2.001 & n\\
 94 & mask 2        &  7 38 17.56 & 38 51 57.5 & -18.071  &  2.288 &  18.286 &   1.036 &  -1.815 &   2.004 & n\\
 95 & mask 1        &  7 38 18.48 & 38 53 32.0 & -20.034  &  2.577 &  19.910 &   0.968 &  -1.820 &   2.036 & y\\
 95 & mask 2        &  7 38 18.48 & 38 53 32.0 & -15.769  &  2.452 &  19.910 &   0.968 &  -1.660 &   2.036 & y\\
 96 & mask 1        &  7 38 18.90 & 38 52 31.8 & -18.255  &  2.267 &  17.237 &   1.204 &  -1.885 &   2.058 & n\\
 97 & mask 2        &  7 38 12.63 & 38 51 01.2 &  48.942  &  2.719 &  19.990 &   0.814 &  -0.742 &   2.058 & n\\
 98 & mask 2        &  7 38 17.64 & 38 51 50.6 & -17.449  &  2.334 &  19.055 &   1.020 &  -1.735 &   2.075 & n\\
 99 & mask 1        &  7 37 58.86 & 38 53 56.7 & -17.789  &  2.334 &  17.790 &   1.077 &  -1.182 &   2.142 & n\\
100 & mask 1        &  7 38 17.81 & 38 54 06.1 &  46.917  &  2.283 &  17.032 &   1.201 &  -1.329 &   2.164 & n\\
101 & mask 2        &  7 37 57.76 & 38 53 29.3 & -16.325  &  2.269 &  16.959 &   1.202 &  -1.473 &   2.169 & n\\
102 & B09  III89    &  7 38 09.79 & 38 50 45.4 & -26.050  &  0.920 &  15.761 &   1.507 &  -3.515 &   2.173 & n\\
103 & B09  III107   &  7 38 15.55 & 38 51 13.5 & -17.430  &  0.990 &  17.156 &   1.230 &  -3.215 &   2.176 & n\\
104 & B09  12       &  7 38 18.63 & 38 53 52.0 & -17.700  &  1.160 &  17.409 &   1.186 &  -3.159 &   2.187 & y\\
104 & mask 2        &  7 38 18.63 & 38 53 52.1 & -19.898  &  2.282 &  17.409 &   1.186 &  -1.679 &   2.188 & y\\
105 & mask 1        &  7 37 57.27 & 38 52 58.0 & -24.719  &  2.331 &  17.816 &   1.122 &  -1.915 &   2.188 & y\\
105 & mask 2        &  7 37 57.27 & 38 52 58.0 & -25.534  &  2.372 &  17.816 &   1.122 &  -1.864 &   2.188 & y\\
106 & mask 1        &  7 38 19.64 & 38 53 15.6 & -28.739  &  2.271 &  15.909 &   1.471 &  -1.774 &   2.193 & n\\
107 & mask 1        &  7 38 18.16 & 38 54 04.2 & -16.560  &  2.557 &  19.591 &   0.982 &  -1.633 &   2.205 & n\\
108 & mask 1        &  7 38 16.77 & 38 51 22.6 & -17.489  &  2.421 &  19.347 &   1.006 &  -1.502 &   2.225 & y\\
108 & mask 2        &  7 38 16.77 & 38 51 22.6 & -20.194  &  2.407 &  19.347 &   1.006 &  -1.481 &   2.225 & y\\
109 & B09  II118    &  7 38 16.93 & 38 54 25.2 & -20.750  &  1.040 &  16.849 &   1.242 &  -3.294 &   2.225 & n\\
110 & mask 1        &  7 38 20.01 & 38 52 48.8 & -26.997  &  2.290 &  17.987 &   1.103 &  -1.522 &   2.240 & n\\
111 & mask 1        &  7 38 17.13 & 38 51 24.5 & -19.471  &  2.383 &  19.084 &   1.021 &  -1.704 &   2.255 & n\\
112 & mask 1        &  7 37 57.88 & 38 51 56.9 & -18.050  &  2.280 &  16.595 &   1.305 &  -1.596 &   2.284 & y\\
112 & mask 2        &  7 37 57.88 & 38 51 56.9 & -17.875  &  2.480 &  16.595 &   1.305 &  -1.570 &   2.284 & y\\
113 & mask 1        &  7 37 57.57 & 38 51 58.9 & -20.987  &  2.392 &  19.232 &   1.017 &  -1.865 &   2.325 & n\\
114 & mask 1        &  7 38 20.54 & 38 53 06.4 & -12.780  &  2.510 &  19.812 &   0.959 &  -1.755 &   2.349 & n\\
115 & mask 2        &  7 38 20.23 & 38 52 12.1 & -19.165  &  2.285 &  17.866 &   1.130 &  -1.776 &   2.390 & n\\
116 & B09  III86    &  7 38 09.54 & 38 50 31.7 & -18.890  &  0.990 &  16.973 &   1.248 &  -3.259 &   2.395 & n\\
117 & mask 1        &  7 37 58.19 & 38 51 36.5 & -16.915  &  2.710 &  19.935 &   0.961 &  -1.725 &   2.396 & n\\
118 & mask 2        &  7 38 18.02 & 38 51 23.1 & -19.906  &  2.279 &  18.138 &   1.103 &  -1.901 &   2.402 & n\\
119 & mask 1        &  7 37 56.40 & 38 53 28.0 & -15.372  &  2.296 &  17.086 &   1.238 &  -1.725 &   2.420 & y\\
119 & mask 2        &  7 37 56.40 & 38 53 28.0 & -15.207  &  2.271 &  17.086 &   1.238 &  -1.726 &   2.420 & y\\
120 & mask 1        &  7 38 15.28 & 38 54 57.7 & -24.280  &  2.747 &  19.578 &   0.982 &  -1.197 &   2.434 & n\\
121 & mask 2        &  7 37 56.68 & 38 53 54.3 & -20.776  &  2.284 &  17.126 &   1.202 &  -1.794 &   2.506 & n\\
122 & mask 1        &  7 38 20.92 & 38 53 44.1 & -21.836  &  2.331 &  18.922 &   1.042 &  -1.851 &   2.550 & n\\
123 & mask 1        &  7 37 55.40 & 38 52 46.0 & -18.686  &  2.436 &  19.246 &   1.037 &  -1.682 &   2.556 & n\\
124 & mask 2        &  7 37 56.48 & 38 51 53.3 & -18.725  &  2.276 &  17.698 &   1.160 &  -1.797 &   2.557 & n\\
125 & mask 1        &  7 37 55.80 & 38 52 13.5 & -21.527  &  2.295 &  17.831 &   1.145 &  -1.753 &   2.568 & y\\
125 & mask 2        &  7 37 55.80 & 38 52 13.5 & -19.102  &  2.291 &  17.831 &   1.145 &  -1.644 &   2.568 & y\\
126 & mask 2        &  7 38 01.20 & 38 50 38.9 & -21.521  &  2.287 &  17.898 &   1.138 &  -1.906 &   2.676 & n\\
127 & mask 1        &  7 37 54.67 & 38 53 00.3 & -20.692  &  2.506 &  19.680 &   1.000 &  -1.711 &   2.695 & n\\
128 & mask 2        &  7 38 08.56 & 38 50 13.0 & -18.569  &  2.453 &  19.612 &   0.901 &  -1.388 &   2.698 & n\\
129 & B09  II59     &  7 38 13.97 & 38 55 26.7 & -23.200  &  1.110 &  17.641 &   1.164 &  -3.102 &   2.744 & n\\
130 & B09  IV93     &  7 37 56.45 & 38 51 22.8 & -20.640  &  0.690 &  16.456 &   1.341 &  -3.373 &   2.805 & n\\
131 & mask 2        &  7 38 22.80 & 38 52 28.1 & -30.504  &  2.269 &  17.466 &   1.177 &  -1.786 &   2.817 & n\\
132 & mask 2        &  7 38 21.96 & 38 51 46.7 & -23.469  &  2.414 &  19.638 &   0.971 &  -1.733 &   2.854 & n\\
133 & mask 1        &  7 38 23.62 & 38 52 30.7 & -21.792  &  2.675 &  19.822 &   0.967 &  -1.652 &   2.968 & y\\
133 & mask 2        &  7 38 23.62 & 38 52 30.7 & -22.507  &  2.615 &  19.822 &   0.967 &  -1.320 &   2.968 & y\\
134 & mask 1        &  7 38 19.20 & 38 50 47.8 & -19.527  &  2.270 &  17.365 &   1.204 &  -1.839 &   2.969 & y\\
134 & mask 2        &  7 38 19.20 & 38 50 47.8 & -19.462  &  2.266 &  17.365 &   1.204 &  -1.814 &   2.969 & y\\
135 & mask 1        &  7 37 54.00 & 38 54 02.5 & -18.755  &  2.347 &  18.664 &   1.057 &  -1.833 &   3.040 & n\\
136 & B09  III37    &  7 38 13.72 & 38 50 00.5 & -21.300  &  1.200 &  17.092 &   1.200 &  -3.240 &   3.079 & n\\
137 & mask 2        &  7 38 14.52 & 38 50 00.2 &  -4.134  &  2.292 &  17.985 &   1.071 &  -1.160 &   3.138 & n\\
138 & mask 2        &  7 38 22.74 & 38 51 26.0 & -20.773  &  2.298 &  18.517 &   0.987 &  -1.935 &   3.141 & n\\
139 & B09  I48      &  7 37 58.06 & 38 55 22.2 & -18.470  &  0.710 &  16.911 &   1.220 &  -3.283 &   3.188 & n\\
140 & B09  I105     &  7 37 54.13 & 38 54 32.7 & -21.430  &  0.970 &  16.892 &   1.255 &  -3.279 &   3.238 & n\\
\tableline
\end{tabular}
\end{center}
\end{table*}

\begin{table*}
\begin{center}
\setcounter{table}{0}
\caption{...continued.}
\begin{tabular}{ccccccccccc}
\tableline
141 & mask 2        &  7 38 08.76 & 38 49 40.2 & -21.026  &  2.503 &  19.655 &   0.980 &  -1.312 &   3.245 & n\\
142 & mask 2        &  7 38 18.47 & 38 50 18.3 & -19.292  &  2.560 &  19.985 &   0.930 &  -1.349 &   3.251 & n\\
143 & mask 1        &  7 37 51.26 & 38 53 07.4 & -22.107  &  2.317 &  18.272 &   1.122 &  -1.837 &   3.363 & n\\
144 & mask 1        &  7 38 23.06 & 38 51 04.5 & -20.391  &  2.322 &  18.527 &   1.033 &  -1.551 &   3.377 & y\\
144 & mask 2        &  7 38 23.06 & 38 51 04.5 & -22.086  &  2.316 &  18.527 &   1.033 &  -1.463 &   3.377 & y\\
145 & mask 1        &  7 38 26.02 & 38 52 42.1 & -20.035  &  2.351 &  18.872 &   1.016 &  -1.596 &   3.414 & n\\
146 & mask 2        &  7 38 18.19 & 38 49 52.9 & -21.514  &  2.314 &  18.866 &   1.021 &  -1.591 &   3.571 & n\\
147 & B09  II23     &  7 38 09.58 & 38 56 30.0 & -20.670  &  0.890 &  16.655 &   1.286 &  -3.334 &   3.591 & n\\
148 & mask 2        &  7 38 23.89 & 38 50 51.3 & -22.526  &  2.270 &  17.190 &   1.211 &  -1.871 &   3.634 & n\\
149 & mask 1        &  7 38 24.59 & 38 54 50.4 & -21.273  &  2.468 &  19.549 &   0.989 &  -1.666 &   3.673 & n\\
150 & mask 1        &  7 37 53.50 & 38 50 35.5 & -31.810  &  2.828 &  19.942 &   0.959 &  -1.598 &   3.733 & n\\
151 & mask 2        &  7 38 26.89 & 38 51 30.7 & -23.012  &  2.380 &  19.129 &   1.020 &  -1.713 &   3.843 & n\\
152 & mask 1        &  7 38 25.26 & 38 55 15.9 &  16.419  &  2.626 &  19.655 &   0.978 &  -0.873 &   4.018 & n\\
153 & mask 1        &  7 37 48.61 & 38 51 20.6 & -22.418  &  2.325 &  18.673 &   0.999 &  -1.610 &   4.180 & n\\
154 & mask 2        &  7 38 27.15 & 38 50 43.3 & -18.856  &  2.522 &  19.915 &   0.962 &  -1.540 &   4.240 & n\\
155 & mask 1        &  7 37 46.06 & 38 51 41.4 & -21.155  &  2.281 &  17.821 &   1.094 &  -1.617 &   4.538 & n\\
156 & mask 2        &  7 38 19.92 & 38 48 50.4 &  18.899  &  2.518 &  19.661 &   0.867 &  -0.709 &   4.641 & n\\
157 & mask 1        &  7 37 45.44 & 38 54 23.6 & -20.425  &  2.356 &  18.950 &   1.015 &  -1.659 &   4.726 & n\\
158 & mask 1        &  7 38 31.48 & 38 51 12.1 & -21.597  &  2.277 &  17.598 &   1.172 &  -1.772 &   4.788 & n\\
159 & mask 1        &  7 38 33.01 & 38 53 45.2 & -21.967  &  2.264 &  16.379 &   1.304 &  -1.452 &   4.841 & n\\
160 & mask 1        &  7 38 32.27 & 38 54 43.4 & -20.964  &  2.292 &  17.018 &   1.233 &  -1.530 &   4.964 & n\\
161 & mask 1        &  7 37 42.49 & 38 51 11.9 & -14.861  &  2.338 &  19.085 &   1.020 &  -1.769 &   5.348 & n\\
162 & mask 1        &  7 37 40.59 & 38 52 03.0 &  13.429  &  2.273 &  16.840 &   1.190 &  -1.277 &   5.502 & n\\
163 & mask 2        &  7 38 31.98 & 38 49 47.7 & -21.679  &  2.429 &  19.249 &   1.006 &  -1.545 &   5.533 & n\\
164 & mask 1        &  7 37 40.96 & 38 51 05.7 & -21.438  &  2.260 &  16.637 &   1.285 &  -1.909 &   5.663 & n\\
165 & mask 2        &  7 38 32.03 & 38 49 13.9 & -52.154  &  2.368 &  18.451 &   1.100 &  -0.993 &   5.877 & n\\
166 & mask 1        &  7 38 41.05 & 38 52 26.0 & -18.550  &  2.322 &  17.867 &   1.053 &  -2.047 &   6.351 & n\\
167 & mask 2        &  7 38 30.47 & 38 47 46.5 & -21.544  &  2.512 &  19.722 &   0.947 &  -1.514 &   6.686 & n\\
168 & mask 1        &  7 37 33.01 & 38 51 39.9 & -25.755  &  2.603 &  19.637 &   0.950 &  -1.705 &   7.022 & n\\
169 & mask 2        &  7 38 42.77 & 38 49 41.0 &-160.538  &  2.735 &  19.925 &   0.828 &  -0.867 &   7.412 & n\\
170 & mask 1        &  7 37 29.30 & 38 52 28.7 & -19.041  &  2.286 &  17.645 &   1.119 &  -1.941 &   7.644 & n\\
171 & mask 2        &  7 38 34.78 & 38 47 04.1 & -21.921  &  2.525 &  19.273 &   0.956 &  -0.822 &   7.769 & n\\
172 & mask 2        &  7 38 48.16 & 38 48 19.6 &  -5.179  &  2.499 &  19.690 &   0.968 &  -1.272 &   8.981 & n\\
173 & mask 1        &  7 37 22.10 & 38 52 45.5 & -32.019  &  2.356 &  18.114 &   1.155 &  -0.977 &   9.033 & n\\
174 & mask 1        &  7 37 22.89 & 38 51 05.9 & -14.079  &  2.506 &  18.711 &   1.120 &  -1.519 &   9.064 & n\\
175 & mask 2        &  7 38 43.77 & 38 46 01.5 & -50.217  &  2.278 &  17.046 &   1.239 &  -1.127 &   9.728 & n\\
176 & mask 2        &  7 38 49.47 & 38 46 22.9 & -49.256  &  2.285 &  17.281 &   1.238 &  -1.082 &  10.311 & n\\
177 & mask 2        &  7 38 55.49 & 38 47 44.3 & -21.062  &  2.540 &  16.729 &   1.317 &  -1.052 &  10.511 & n\\
178 & mask 2        &  7 38 56.74 & 38 43 54.3 & -22.239  &  2.533 &  19.340 &   0.876 &  -1.066 &  13.018 & n\\
\tableline\tableline
\end{tabular}
\end{center}
\end{table*}

\end{document}